\documentclass[onecolumn,tighten,times]{aastex6}
\usepackage{color}
\usepackage{graphicx}
\usepackage{mathrsfs}
\usepackage{ulem}
\usepackage{empheq}
\usepackage{multirow}
\usepackage{array}

\DeclareMathAlphabet{\mathbi}{OT1}{ptm}{bx}{it}
\SetMathAlphabet\mathbi{bold}{OT1}{ptm}{bx}{it}

\def\sss{\scriptscriptstyle}
\def\bhm{M_{\bullet}}

\def\ergs{\rm erg~s^{-1}}
\def\fblr{f_{\sss\rm BLR}}
\def\msun{M_{\odot}}
\def\mathdotM{\dot{\mathscr{M}}}
\def\rblr{R_{\sss{\rm BLR}}}
\def\sunm{M_\odot}
\def\tauhb{\tau_{_{\rm H\beta}}}
\def\taud{\tau_{\rm d}}

\def\mathdotM{\dot{\mathscr{M}}}

\begin{document}
\title{Supermassive black holes with high accretion rates in active galactic nuclei: 
X. Optical variability characteristics}

\author{Kai-Xing Lu\altaffilmark{1,2,9},
Ying-Ke Huang\altaffilmark{3},
Zhi-Xiang Zhang\altaffilmark{3},
Kai Wang\altaffilmark{3},
Pu Du\altaffilmark{3},
Chen Hu\altaffilmark{3},
Ming Xiao\altaffilmark{3},
Yan-Rong Li\altaffilmark{3},
Jin-Ming Bai\altaffilmark{1,2,9}, 
Wei-Hao Bian\altaffilmark{4},
Ye-Fei Yuan\altaffilmark{5},
Luis C. Ho\altaffilmark{6,7} and
Jian-Min Wang\altaffilmark{3,8,9}\\
(SEAMBH collaboration)}

\altaffiltext{1}{Yunnan Observatories, Chinese Academy of Sciences, Kunming 650011, Yunnan, China}
\altaffiltext{2}{Key Laboratory for the Structure and Evolution of Celestial Objects, Chinese Academy of Sciences, Kunming 650011, China}
\altaffiltext{3}{Key Laboratory for Particle Astrophysics, Institute of High Energy Physics, Chinese Academy of Sciences, 19B Yuquan Road, Beijing 100049, China.}
\altaffiltext{4}{Physics Department, Nanjing Normal University, Nanjing 210097, China}
\altaffiltext{5}
{Department of Astronomy, University of Science and Technology of China, Hefei 230026, China}
\altaffiltext{6}{Kavli Institute for Astronomy and Astrophysics, Peking University, Beijing 100871, China}
\altaffiltext{7}{Department of Astronomy, School of Physics, Peking University, Beijing 100871, China}
%\altaffiltext{8}{School of Astronomy and Space Science, University of Chinese Academy of Sciences, 19A Yuquan Road, Beijing 100049, China}
\altaffiltext{8}{National Astronomical Observatories of China, The Chinese Academy of Sciences, 20A Datun Road, Beijing 100020, China}
\altaffiltext{9}{Corresponding authors: lukx@ynao.ac.cn, baijinming@ynao.ac.cn, wangjm@ihep.ac.cn}

\begin{abstract} 
We compiled a sample of 73 active galactic nuclei (AGNs) with reverberation mapping (RM) observations 
from RM campaigns including our ongoing campaign of monitoring 
super-Eddington accreting massive black holes (SEAMBHs). This sample covers a large range of 
black hole (BH) mass $(M_{\bullet}=10^{6-9}~M_\odot)$, dimensionless accretion rates $(\dot{\mathscr{M}}=10^{-2.7}-10^{2.7})$ and 
5100~\AA~luminosity $(L_{5100}=10^{42-46}~\rm erg~s^{-1})$, allowing us to systematically study the 
AGN variability and their relations with BH mass, accretion rates, and optical luminosity.
We employed the damped random walk (DRW) model to delineate the optical 
variability of continuum at 5100~\AA~and obtained damped variability timescale ($\tau_{\rm d}$) and amplitude 
($\sigma_{\rm d}$) using a Markov Chain Monte Carlo (MCMC) method. 
We also estimated the traditional variability amplitudes ($F_{\rm var}$),
which provide a model-independent measure and therefore are used to test the DRW results. 
We found that AGN variability characteristics are generally correlated with
$(M_{\bullet},\dot{\mathscr{M}},L_{5100})$. These correlations are smooth from sub-Eddington 
to super-Eddington accretion AGNs, probably implying that the AGN variability may be caused by the 
same physical mechanism.  
\end{abstract}


\section{Introduction}
\label{sec:intr}
Accretion onto supermassive black holes (BHs) is commonly believed to be the powerful energy 
source of active galactic nuclei (AGNs; e.g., \citealt{Rees1984}), which is evidenced by the 
prominent big blue bumps in AGN spectrum energy distributions 
(e.g. \citealt{Shields1978};
\citealt{Malkan1983};\citealt{Ho2008}). However, the detailed physics of accretion disks are still insufficiently 
understood, even for the basic process of energy dissipation (\citealt{Lawrence2018}). 
AGNs have long been known to show aperiodic variability across a broad wavelength band from radio to X-ray
at various time scales, which delivers useful information on emissions from accretion disk (\citealt{Ulrich1997}). 
Many previous works studied optical variability of AGNs based on various samples 
\citep{Giveon1999,MacLeod2010,Zuo2012,Lu2016b} and made great efforts to investigate 
the relationships between variability characteristics and AGN properties, such as, optical luminosity, BH mass, 
and Eddington ratio (e.g., \citealt{Sanchez2018,Rakshit2017,Wold2007,Wilhite2008,Bauer2009,Zuo2012,Hook1994,
Cristiani1997,Giveon1999,Hawkins1996,Vanden2004}). 
 
Spectroscopic monitoring campaigns have successfully probed 
geometry and kinematics of the broad-line region (BLR), but also have accumulated precious variability databases. 
The major advantage of RM databases is that the AGN properties, such as BH mass and accretion rate, 
can be more reliably estimated. In addition, the quality of light curves from RM campaigns (e.g., sampling and 
measurement accuracy)
is generally better than that from other time-domain surveys (except for some light curves observed by the \textit{Kepler} telescope). 
In 2012, we started a long-term RM campaign aiming at monitoring super-Eddington accreting massive black holes 
(SEAMBHs) using the Lijiang 2.4m telescope (\citealt{Du2014}; \citealt{Du2018}). 
By combining with RM AGNs from previous RM campaigns (e.g., \citealt{Bentz2013}), there are $\sim$100 AGNs 
monitored by RM campaigns\footnote{Besides our new SEAMBH targets, several new RM objects published during our 
SEAMBH campaign period are also included. MCG-06-30-15 from \cite{Bentz2016a}; UGC 06728 from \cite{Bentz2016b}; 
NGC 5548 from \cite{Lu2016a}; MCG+08-11-011, NGC 2617, NGC 4051, 3C 382, and Mrk 374 from \cite{Fausnaugh2017}.
}. 
This sample provides with us a good opportunity to study AGN variability characteristics, and their connections with AGN basic 
properties (BH mass, accretion rates, and optical luminosity).

The paper is organized as follows. 
We describe the sample and AGN properties in Section~\ref{sec:sample}. 
Section~\ref{sec:method} presents the methodology and variability characteristics of AGNs. 
We test the validity of the DRW model in Section~\ref{sec:valid}, 
and investigate the differences of variability characteristics between 
super- and sub-Eddington accretion AGNs in Section~\ref{sec:diff}. 
Section~\ref{sec:cor} performs a correlation analysis between variability characteristics and AGN properties. 
We draw our conclusion in Section~\ref{sec:con}. 

\section{Sample and Properties}
\label{sec:sample}
For the purpose of accurately determining the BH mass and accretion rates, 
we only selected AGNs with RM observations that detect significant H$\beta$ time delays. 
Over the past 30 years, great efforts have been made to conduct RM monitoring of nearby AGNs 
(e.g., \citealt{Peterson1998, Kaspi2000, Bentz2013, Denney2009, Grier2012}).
\citet{Du2015} compiled all the published RM measurements before the year of 2014 (see Table 7 
of \citealt{Du2015}), and most of the mapped AGNs before 2013 are sub-Eddington AGNs. 
Recently, the SDSS-RM project monitored a sample of AGNs with redshift up 
to $z\sim0.3$ (\citealt{Shen2015, Grier2017}), but their database is not available. 
The SEAMBH project provided a complementary sample of 
super-Eddington accretion AGNs, and the latest SEAMBH results are reported in paper 
of \citet{Du2016,Du2018}. We finally obtain a sample of 73 AGNs with 113 light curves (because some objects have multiple 
RM campaigns, such as NGC 5548 and Mrk 335). 
The light curves do not include the component of the broad-emission line since 
they were measured from the optical spectra at 5100~\AA~
(for references see column 11 of Table~\ref{table:drw}). 
For these objects observed with multiple campaigns, we do not combine their 
measurements but instead treat the measurements independently in the following analysis. 
It should be noted that AGN luminosity at 5100~\AA~is contaminated by emission from stars in the host galaxy, 
especially, the host to optical luminosity ratio ($L_{\rm host}/L_{\rm opt}$) maybe larger than 50\% in low-luminosity AGNs 
(see Figure 13 of \citealt{Stern2012}). In our sample, 
the optical luminosity $L_{5100}$ has been corrected for the starlight of the host galaxy 
either by HST data or empirical relation (see \citealt{Bentz2013,Du2015,Du2018}), 
in which the characteristic ratio of host to optical luminosity is $L_{\rm host}/L_{\rm opt}=0.41$ 
with a scatter of 0.21. This larger scatter is consistent with the fact that 
the host to AGN luminosity (e.g., $L_{\rm 5100}$) ratio is a function of AGN bolometric luminosity \citep{Shen2011,Stern2012}. 
Table~\ref{table:drw} summarizes the basic properties of the sample.

With the usual assumption that motion of the BLR clouds is virialized, 
we estimate BH mass by 
\begin{equation}
\bhm=\fblr \frac{V_{\rm BLR}^2\rblr}{G},
\end{equation}
where $\rblr=c\tauhb$, $\tauhb$ is the H$\beta$ time lag with respective to the 5100~{\AA} continuum, 
$G$ is the gravitational constant, $c$ is the speed of light, and $\fblr$ is the so-called virial factor that
includes all the unknown information about the geometry and kinematics of the BLR gas. 
The dimensionless accretion rate is related to the 5100 \AA\, luminosity and BH mass via
\citep{Du2014}
\begin{equation}
\mathdotM=\frac{\dot{M}_{\bullet}}{L_{\rm Edd}c^{-2}}=20.1\left(\frac{\ell_{44}}{\cos i}\right)^{3/2}M_7^{-2},
\label{eqn_mdot}
\end{equation}
where $\dot{M}_{\bullet}$ is the accretion rates, $\ell_{44}=L_{5100}/10^{44}\ergs$, 
$M_7=\bhm/10^7\sunm$ is the BH mass, and $\cos i$ is the cosine of the inclination of the accretion disk. 
Following the usual approximation, we take $\cos i=0.75$, corresponding to an inclination angle of $\sim 40^\circ$.

\begin{figure*}[b]
\centering
\includegraphics[angle=0,width=0.99\textwidth]{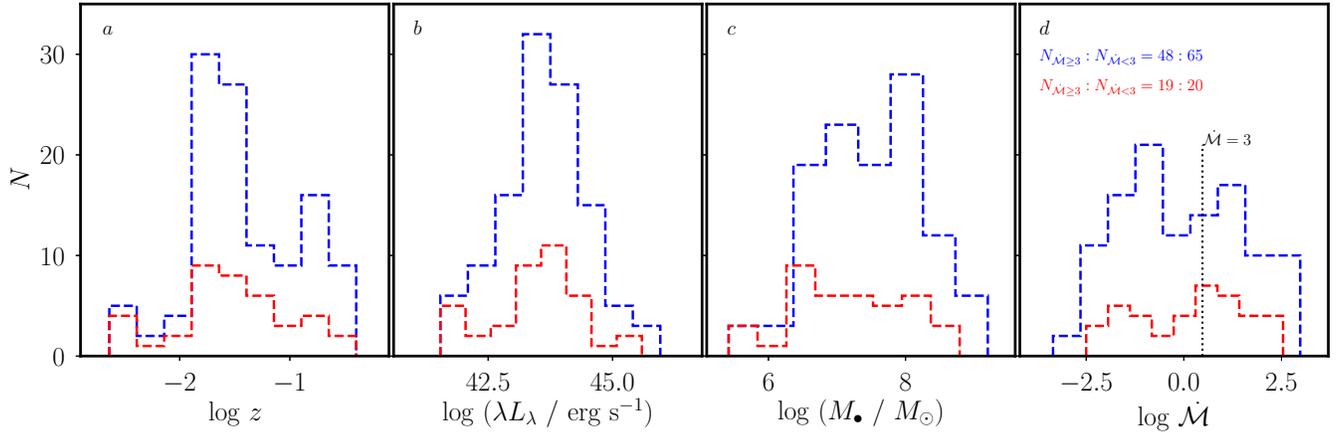}
\caption{\footnotesize 
The distributions of AGN properties: (a) redshift, (b) optical luminosity, (c) black hole mass,
and (d) accretion rates. Blue dashed-line is for all RM AGNs 
and red dashed-line is for the AGNs with  $\tau_{\rm d}|_{<0.1D}$ (see Section~\ref{sec:valid}). 
}
\label{fig:agnp}
\end{figure*}

Figure~\ref{fig:agnp} shows the distributions of redshifts ($z$), 5100~{\AA} luminosity ($L_{\rm 5100}$), 
BH mass 
($\bhm$), and dimensionless accretion rates ($\mathdotM$) of our sample. 
Following \cite{Du2015}, we classify AGNs into sub-Eddington and super-Eddington regimes by $\mathdotM=3$, 
beyond which the inner parts of disk break the standard model of accretion disk \citep{Laor1989}, 
the radial advection of accrete flows is not negligible. 
The AGNs with high accretion rates ($\mathdotM\gtrsim 3$) are thought to be powered by slim disk \citep{Abramowicz1988}. 
We have a number ratio of the two subsamples $N_{\mathdotM\geq3}:N_{\mathdotM<3}=48:65$, 
implying that the entire sample covers  a homogeneous range of accretion rates. 

Here we would like to point out the difference between $\mathdotM$ and Eddington ratios defined by
$\lambda_{\rm Edd}=L_{\rm Bol}/L_{\rm Edd}$, where $L_{\rm Bol}$ 
is the bolometric luminosity. In the Shakura-Sunyaev regime \citep{Shakura1973}, we have
$L_{\rm Bol}=\eta\dot{M}_{\bullet}c^2$, 
where $\eta$ is the radiative efficiency depending on BH spins. We have 
$\lambda_{\rm Edd}=\eta\mathdotM$ from 
Equation (\ref{eqn_mdot}), that is $\mathdotM$ is linearly proportional to $\lambda_{\rm Edd}$, 
indicating that $\mathdotM$ and $\lambda_{\rm Edd}$ represent the accretion rates in sub-Eddington accreting AGNs. 
However, $\lambda_{\rm Edd}$ cannot be an indicator of accretion rates in super-Eddington AGNs. 
Beyond the Shakura-Sunyaev model, the slim accretion disk is characterized by its fast 
radial motion compared with Keplerian rotation giving rise to nonlocal energy budget. In this case, 
photons produced in the viscose dissipation are trapped inside accretion flow so that a large fraction of 
photons are swallowed into the BH before they escape from the disk surface. 
This photon trapping effect leads to that the radiated luminosity 
($L_{\rm Bol}$) being saturated  when $\mathdotM\gg1$ (see Figure 1 of \citealt{Abramowicz1988} or 
Figure 2 of \citealt{Mineshige2000}). 
In such a case, $\lambda_{\rm Edd}\sim 1$, but $\mathdotM\gg1$, 
indicating that Eddington ratios cannot represent accretion rates in super-Eddington AGNs.
Earlier discussions on the validity of the Shakura-Sunyaev disk show that geometrically thin approximation is 
broken for $\mathdotM\gtrsim 3$ \citep[see details in][]{Laor1989} extending to the regime of slim 
disk \citep{Wang2014}. We thus take this approach in this paper.

Moreover, applying the $R-L$ relation and its lines extended to Mg {\sc ii} and C {\sc iv}, astronomers
estimated Eddington ratios of a 
large sample, such as the Sloan Digital Sky Survey (SDSS) \citep[e.g.,][]{McLure2002,Shen2011},
the AGN and Galaxy Evolution Survey (AGES) \citep{Kollmeier2006}, where the bolometric correction factor
is usually taken to be about 10. This factor is too small for some AGNs \citep{Jin2012}.
They found $\lambda_{\rm Edd}\lesssim 1$
in quasars. The long-term ongoing SEAMBH campaign 
is getting more evidence for shortened H$\beta$ lags for optical Fe {\sc ii} strong 
AGNs \citep{Du2015,Du2016,Du2018}, implying that BH masses in some quasars are overestimated and 
Eddington ratios are underestimated. Estimations of BH mass and Eddington ratios need to be improved.

\section{Methodology and Variability Characteristics}
\label{sec:method}
\subsection{Damped Random Walk Model}
Damped random walk (DRW) is a stochastic process, 
defined by an exponential covariance function (e.g., \citealt{Zu2011, Zu2013}), 
\begin{equation}
S(t_1, t_2)=\sigma_{\rm d}^2\exp\left[-\left(\frac{|t_i-t_j|}{\tau_{\rm d}}\right)\right],
\label{eqn_cov}
\end{equation}
where $\tau_{\rm d}$ is the characteristic variability timescale in units of days, 
$t_i-t_j=\Delta t$ is the time sampling interval between the $i$th and $j$th observations, and 
$\sigma_{\rm d}$ is the characteristic variability amplitude in units of flux density. 
It has been shown that AGNs variability in the optical band can be well described by the DRW (e.g., \citealt{Kelly2009}; 
\citealt{Kozlowski2010}; \citealt{MacLeod2010, MacLeod2011}; \citealt{Zu2013};\citealt{Rakshit2017}). 
The model has been widely used in modeling AGN light curves in 
reverberation mapping (RM) studies (see \citealt{Zu2011,Pancoast2011,Grier2012,LiYanRong2013,Zhang2018}). 
The limitations of the DRW model in describing AGN variability characteristics will be discussed in Section~\ref{sec:valid}. 

The framework for estimating DRW model parameters and reconstructing AGN light curves has been 
well-established, e.g, see  \cite{Rybicki1992} and \cite{Zu2011}. Here we briefly list the 
essential points for the sake of completeness. Following \citealt{Rybicki1992}, we model light curves as 
\begin{equation}
\mathbi{y} = \mathbi{s+n}+\mathbi{E}\mathbi{q}, 
\end{equation}
where $\mathbi{s}$ is the underlying variability signal with covariance matrix $\mathbi{S}$, 
$\mathbi{n}$ is the measurement errors with covariance matrix $\mathbi{N}$, 
$\mathbi{E}$ is a vector with all unity elements (i.e. $E_i=1$), and $\mathbi{q}$ is the mean 
value in the light curves. The overall covariance matrix of data is $\mathbi{C}=\mathbi{S}+\mathbi{N}$. 
The posterior probability of the observed data for a given set of DRW model parameters is \citep{Zu2011,Kozlowski2010}
\begin{eqnarray} 
P(\mathbi{y}|\sigma_{\rm d}, \tau_d)\propto \mathcal{L} \equiv 
|\mathbi{C}|^{-1/2}|\mathbi{E}^{T}\mathbi{C}^{-1}\mathbi{E}|^{-1/2}
\exp\left(-\frac{1}{2}\mathbi{y}^T\mathbi{C}_\perp^{-1}\mathbi{y}\right),
\label{eqn_post_var}
\end{eqnarray}
where 
\begin{eqnarray} 
\mathbi{C}_\perp^{-1}=\mathbi{C}^{-1}-\mathbi{C}^{-1}\mathbi{E} (\mathbi{E}^{T} \mathbi{C}^{-1} \mathbi{E}) \mathbi{E}^{T}\mathbi{C}^{-1}. 
\end{eqnarray}
We maximize the posterior probability to determine the best estimate of $\sigma_{\rm d}$ and $\tau_d$ and their uncertainties. 

After determining the values of $\sigma_{\rm d}$ and $\tau_{\rm d}$, 
an unbiased estimate of the light curve at time $t_{\ast}$ is \citep{Rybicki1992}
\begin{eqnarray} 
\widehat{y_{\ast}} = \mathbi{S}_{\ast}^{T}\mathbi{C}^{-1}\mathbi{(y-E}\hat{q}) + \hat{q}. 
\label{eqn_recon}
\end{eqnarray}
The mean square residual of this estimate is \citep{LiYanRong2013} 
\begin{eqnarray} 
\langle (y_{\ast}-\widehat{y_{\ast}})^{2}\rangle=\langle y_{\ast}^{2}\rangle - \mathbi{S}_{\ast}^{T}\mathbi{C}^{-1}\mathbi{S_{\ast}} + 
\frac{(\mathbi{S}_{\ast}^{T}\mathbi{C}^{-1}\mathbi{E}-1)^{2}} {\mathbi{E}^{T}\mathbi{C}^{-1}\mathbi{E}}. 
\label{eqn_recon_err}
\end{eqnarray}

\subsection{Intrinsic Variability Amplitude}
\label{sec_TVC}
The widely used variability amplitude of an AGN light curve is defined as (\citealt{Rodriguez1997})  
\begin{equation}
F_{\rm var}=\frac{\left(\sigma^2-\Delta^2\right)^{1/2}}{\langle F\rangle} \,.
\end{equation}
The uncertainty of $F_{\rm var}$ is defined as (see \citealt{Edelson2002}), 
\begin{equation}
\sigma_{_{F_{\rm var}}} = \frac {1} {F_{\rm var}} \left(\frac {1}{2 N}\right)^{1/2} \frac {\sigma^2}{\langle F\rangle},
\quad \sigma^2=\frac{1}{N-1}\sum_{i=1}^N\left(F_i-\langle F\rangle\right)^2, 
\end{equation}
where $\langle F\rangle=N^{-1}\sum_{i=1}^NF_i$ is the average flux, $N$ is the number of observations, 
$\Delta^2=\sum_{i=1}^N\Delta_i^2/N$, 
$\Delta_i$ represents the uncertainty on the flux $F_i$. 
$F_{\rm var}$ gives an estimate of the relative intrinsic variability amplitude by accounting for the measurement uncertainties.

\begin{figure*}[t!]
\centering
\includegraphics[angle=0,width=1.0\textwidth]{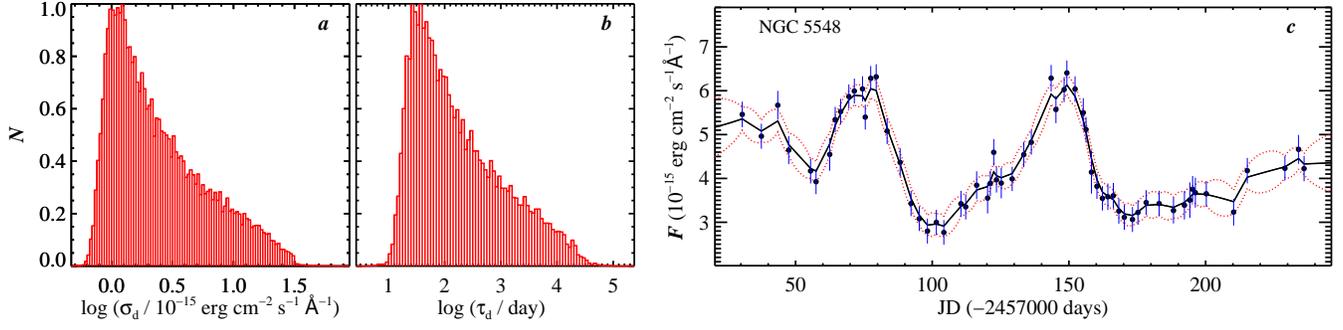}
\caption{\footnotesize
Estimates for the DRW model parameters and a reconstruction of light curve for NGC 5548 from \cite{Lu2016a}. 
Panels ({\it a, b}) show the posterior distributions of $\sigma_{\rm d}$ and $\tau_{\rm d}$ constructed from an MCMC 
method. The distributions were normalised by the peak values. 
In Panel ({\it c}), the points with error bars represent the observed data. 
The solid line represents the best reconstruction and the dashed lines represent the 1$\sigma$ uncertainties.
}
\label{fig:rec}
\end{figure*}

\subsection{Variability Characteristics}
To estimate the DRW model parameters $\sigma_{\rm d}$ and $\tau_{\rm d}$ 
in Equation~(\ref{eqn_cov}), we employed the Markov Chain Monte Carlo (MCMC) method and the 
Metropolis-Hastings algorithm to construct a sample from the posterior probability distribution. 
By maximizing the posterior probability distribution (Equation~\ref{eqn_post_var}), 
we obtain the best estimates of $\sigma_{\rm d}$ and $\tau_{\rm d}$. 
In Figure~\ref{fig:rec}, we take a light curve of NGC 5548 as an example and apply the DRW model. 
Panels ({\it a}) and ({\it b}) of Figure~\ref{fig:rec} show the 
posterior distributions of $\sigma_{\rm d}$ and ($\tau_{\rm d}$), respectively. 
We take the best estimates of $\sigma_{\rm d}$ and $\tau_{\rm d}$ 
to be the peak location of the distributions and the uncertainties to be the 68.3\% confidence interval. 
The underlying signal of light curves can be reconstructed using the best estimates 
of $\sigma_{\rm d}$ and $\tau_{\rm d}$. Panel ({\it c}) of Figure~\ref{fig:rec} shows the best 
reconstruction (solid line) of the light curve of NGC~5548. In the next analysis, 
we normalized $\sigma_{\rm d}$ using average flux of light curve and called it as 
$\Sigma_{\rm d}$ (i.e. $\frac {\sigma_{\rm d}}{\langle F\rangle}$, where $\langle F\rangle$ is average flux of light curves). 
The values of $\Sigma_{\rm d}$, $\tau_{\rm d}$, $F_{\rm var}$ and $R_{\rm max}$ for all RM AGNs 
are tabulated in Table~\ref{table:drw}. 

\section{Test Validity of DRW Parameters}
\label{sec:valid}
\begin{figure}%[ht!]
\centering
\includegraphics[angle=0,width=0.49\textwidth]{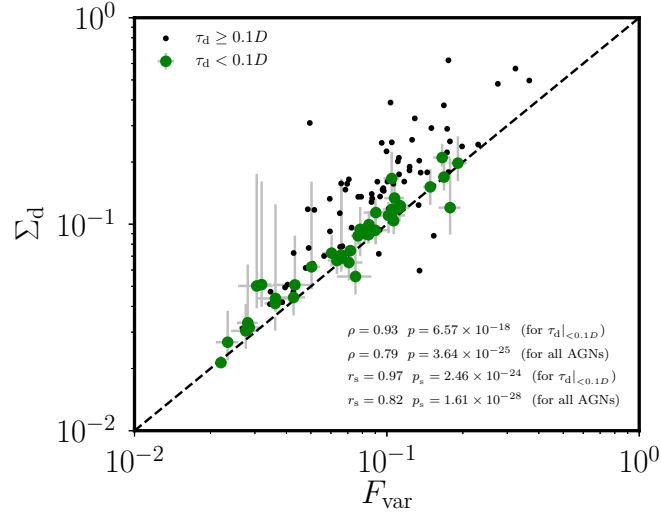}
\caption{\footnotesize
Comparison between the damped variability amplitude and intrinsic variability amplitude. 
The number ratio of $\tau_{\rm d}|_{<0.1D}$ AGNs (in green) and 
$\tau_{\rm d}|_{\ge0.1D}$ AGNs (in black) is 39:74. The dashed line presents $\Sigma_{\rm d}=F_{\rm var}$. 
($\rho, p$) and ($r_{\rm s}, p_{\rm s}$) are the results of the Pearson correlation and Spearman's rank correlation analysis, respectively. 
}
\label{fig:valid}
\end{figure}

The previous studies based on  \textit{Kepler} observations found that the AGN light curves may
deviate from the DRW model and concluded that AGN variability is so complex that more 
sophisticated models are required \citep{Mushotzky2011,Kasliwal2015}. 
To explore possible deviations, \cite{Zu2013} considered a set of AGN variability models 
and found that the light curves of the OGLE quasar are well described by the DRW model. 
Studying the influences of light-curve length, magnitude, and cadence on recovering the DRW model parameters, 
\cite{Kozlowski2017} found that the light curves without enough time length cannot reliably constrain damped variability timescale, 
and suggested that the time length of light curves must be at least 10 times longer than the true damped variability timescale. 
Based on this point, we divided all RM AGNs into two subsamples, they are 
(1) the $\tau_{\rm d}|_{<0.1D}$ sample in which the damped variability timescales are less than 10\% of the observation length of light curves 
(including 39 measurements, where `$D$' means the observation length of light curves), 
and (2) the $\tau_{\rm d}|_{\ge0.1D}$ sample in which the damped variability timescales are greater than or equal to 10\% of the observation length of light curves (including 74 measurements).
After comparing the damped variability amplitude ($\Sigma_{\rm d}$, model-dependent) 
with intrinsic variability amplitude ($F_{\rm var}$, model-independent), 
we found that $\Sigma_{\rm d}$ is approximately equal to $F_{\rm var}$ in the $\tau_{\rm d}|_{<0.1D}$ sample (see Figure~\ref{fig:valid}), 
but deviate from $F_{\rm var}$ in the $\tau_{\rm d}|_{>0.1D}$ sample. 

We employed the Pearson correlation and Spearman's rank correlation analysis to quantitatively compare the relation 
between $\Sigma_{\rm d}$ and $F_{\rm var}$ of two subsamples, the results were shown in Figure~\ref{fig:valid}. 
We found that $\Sigma_{\rm d}$ is linearly correlated with $F_{\rm var}$ in the $\tau_{\rm d}|_{<0.1D}$ sample. 
However, if we included the $\tau_{\rm d}|_{\ge0.1D}$ sample (plotted in black dots), 
$\Sigma_{\rm d}$ gradually deviates from $F_{\rm var}$. 
We further performed a linear regression using the {\tt LinMix} method 
(\citealt{Kelly2007})\footnote{https://github.com/jmeyers314/linmix} for $\tau_{\rm d}|_{<0.1D}$ sample. 
The regression method assumes: $y=\alpha+\beta x +\epsilon$, where $\epsilon$ is the intrinsic scatter about the regression. 
The regression analysis gives
\begin{equation}
\log \Sigma_{\rm d}=(0.09 \pm 0.07) + (1.05\pm 0.06)\log F_{\rm var}~~({\rm for}~\tau_{\rm d}|_{<0.1D}), 
\end{equation}
with intrinsic scatter of $\epsilon=0.03$. 
These checks confirm the suggestion that the DRW model can return reliable DRW model parameters 
if the observation length of light curves is 10 times longer than the true damped variability timescale 
(i.e. $D>10~\tau_{\rm d}$, see \citealt{Kozlowski2017}). 
Therefore, to obtain reasonable results, we only used  the AGNs with $\tau_{\rm d}<0.1D$ 
to investigate the variability characteristics in the following analysis. 
In the statistical analysis of variability amplitudes $F_{\rm var}$ and $R_{\rm max}$, 
we included the objects with $\tau_{\rm d}\ge0.1D$ in the sample because they are model-independent variability amplitudes. 
The distributions of AGN physical properties 
($z$, $L_{5100}$, $\bhm$, and $\mathdotM$) 
of the $\tau_{\rm d}|_{<0.1D}$ sample are plotted (red dashed line) in Figure~\ref{fig:agnp}, 
which gives a number ratio $N_{\mathdotM\geq3}:N_{\mathdotM<3}=19:20$. 

In addition, based on 1384 variable AGNs from the QUEST-La Silla AGN variability survey, 
\cite{Sanchez2018} found that the DRW variability amplitude is affected by the length of the light curve, 
and concluded that the variability of 74\% of AGNs can be described by the DRW model in their defined sample. 
While, our analysis shows that only 35\% (39$/$113) of light curves can be obtained with reliable DRW 
model parameters in RM AGNs. This percentage is significant lower than \cite{Sanchez2018} results, 
which may be attributed to the fact that the light curves length ($\sim$ 1 yr) of 
the RM AGNs is shorter than \cite{Sanchez2018} sample. 

\begin{figure}%[ht!]
\centering
\includegraphics[angle=0,width=0.5\textwidth]{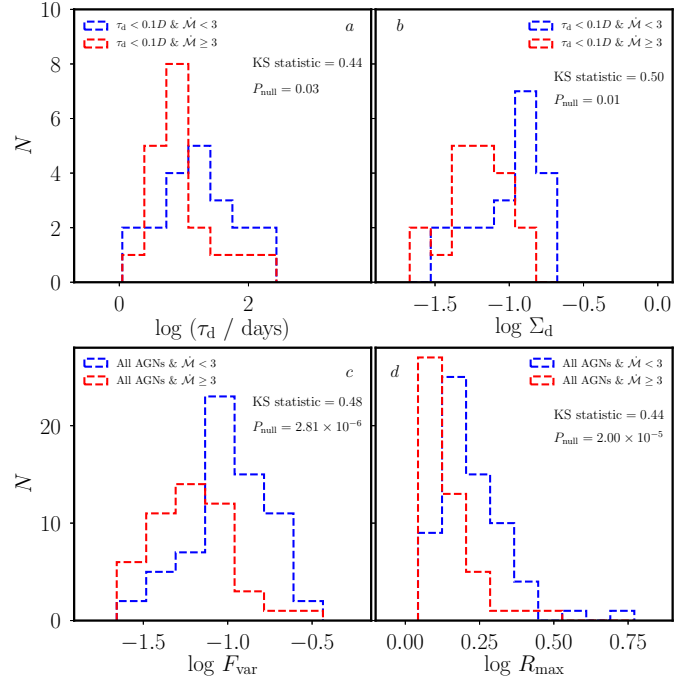}
\caption{\footnotesize
Distributions of variability characteristics. 
Panels ({\it a}) and ({\it b}) are DRW model parameters of $\tau_{\rm d}|_{<0.1D}$ sample. 
Panels ({\it c}) and ({\it d}) are intrinsic variability amplitude and flux ratio, 
which include all RM AGNs ($\tau_{\rm d}|_{<0.1D}$ and $\tau_{\rm d}|_{\ge0.1D}$) since both parameters are model-independent. 
}
\label{fig:pd}
\end{figure}

\section{Differences between Sub- and Super-Eddington Accreting AGNs}
\label{sec:diff}
In this section, we investigate the differences of variability characteristics between sub-Eddington and super-Eddington accreting 
AGNs. 
Figure~\ref{fig:pd} plots the distributions of variability characteristics $\tau_{\rm d}$, $\Sigma_{\rm d}$, $F_{\rm var}$ and $R_{\rm max}$. 
To quantify the differences of variability characteristics between $\mathdotM\geq3$ and $\mathdotM<3$ subsamples, 
we employed the Kolmogorov-Smirnov (KS) statistic test. The results were quoted in Figure~\ref{fig:pd}. 
For DRW model parameters, we only plotted the distributions of $\tau_{\rm d}|_{<0.1D}$ AGNs. 
The KS statistic tests show that the distributions of DRW model parameters 
for $\mathdotM\geq3$ and $\mathdotM<3$ subsamples are marginally different. 
For the model-independent variability parameters ($F_{\rm var}$ and $R_{\rm max}$), 
we plotted the distributions of all RM AGNs because they are model-independent. 
The KS statistic and the $P_{\rm null}$ show that the distributions of both $F_{\rm var}$ and $R_{\rm max}$ 
for $\mathdotM\geq3$ and $\mathdotM<3$ subsamples 
are different. These results indicate that the $\mathdotM\geq3$ AGNs have lower variability amplitude 
($F_{\rm var}$, $R_{\rm max}$ as well as $\Sigma_{\rm d}$) than the $\mathdotM<3$ AGNs, 
which supports the previous notion that AGNs with high accretion rates 
have systematically low variability (e.g.,  \citealt{Wilhite2008,Bauer2009,Zuo2012,Rakshit2017}). 

\begin{figure*}[ht!]
\centering
\includegraphics[angle=0,width=1.0\textwidth]{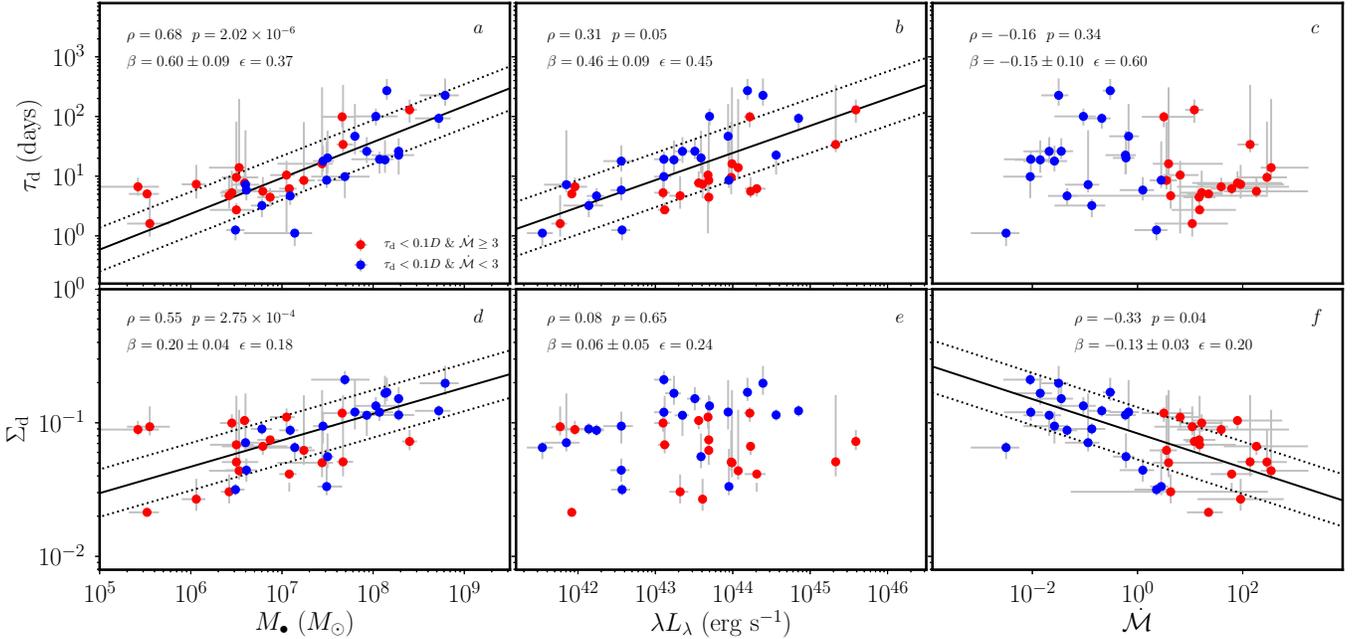}
\caption{\footnotesize
Dependence of DRW model parameters on AGN properties for $\tau_{\rm d}|_{<0.1D}$ sample. 
The top ({\it a, b, c})  and bottom ({\it d, e, f})  panels 
show the damped variability timescale and amplitude as a function of black hole mass, 
optical luminosity and the accretion rates, respectively. 
Solid lines are the best fit and dotted-lines show the intrinsic scatters. 
Red points are the AGNs with $\tau_{\rm d}<0.1D$ and $\mathdotM\geq3$, 
blue points are the AGNs with $\tau_{\rm d}<0.1D$ and $\mathdotM<3$. 
The Pearson correlation coefficient $\rho$ and non-correlation probability $p$ are quoted. 
$\beta$ and $\epsilon$ are the slope and intrinsic scatter, respectively. 
}
\label{fig:dp-ap}
\end{figure*}

\section{Correlation Analysis}
\label{sec:cor}
\subsection{Variability Characteristics and AGN Properties}
\begin{figure*}%[htb!]
\centering
\includegraphics[angle=0,width=1.00\textwidth]{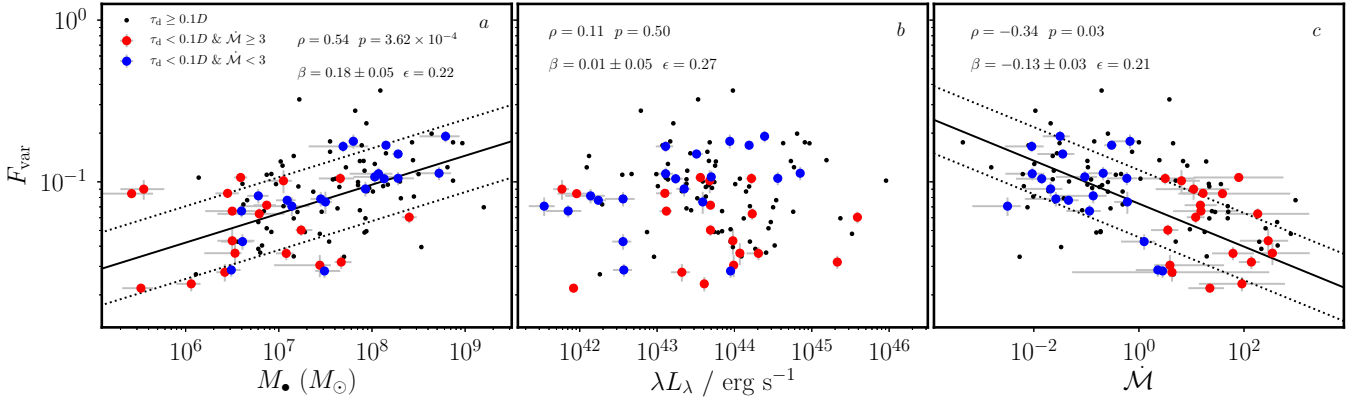}
\caption{\footnotesize 
The same as Fig.~\ref{fig:dp-ap}, but for the relation between intrinsic variability amplitude and AGN properties. 
In panels, we included $\tau_{\rm d}|_{\ge0.1D}$ sample (in black dot) since $F_{\rm var}$ is a model-independent variability amplitude, 
but only quoted the analysis results ($\rho$, $p$) and ($\beta$, $\epsilon$) of $\tau_{\rm d}|_{<0.1D}$ sample. 
We listed the correlation analysis results of all RM AGNs into Table~\ref{tab:pr}, 
and wrote the regression results into Equation~\ref{eqn_reg3}. 
We also only showed the best fit (solid lines) and intrinsic scatters (dotted-lines) of $\tau_{\rm d}|_{<0.1D}$ sample.  
}
\label{fig:var-ap}
\end{figure*}

In Figures~\ref{fig:dp-ap} and~\ref{fig:var-ap}, we plotted the model-dependent variability parameters and the 
intrinsic variability amplitude as a function of AGN properties, respectively. 
Then we calculated the Pearson correlation coefficient $\rho$ and 
noncorrelation probability $p$ between variability characteristics and AGN properties. 
All results are listed in Table~\ref{tab:pr}. 
We also performed a linear regression using the {\tt LinMix} method. 
The results of the linear regression for $\tau_{\rm d}$ are
\begin{subequations}
\begin{empheq}[left={\log \tau_{\rm d}=\empheqlbrace}]{align} 
 &(-3.24 \pm 0.68) + (0.60 \pm 0.09) \log \bhm~~({\rm for}~\tau_{\rm d}|_{<0.1D}), \\
 &(-18.52 \pm 4.06) + (0.46 \pm 0.09) \log L_{5100}~~({\rm for}~\tau_{\rm d}|_{<0.1D}), 
% &(1.26 \pm 0.12) - (0.10 \pm 0.10) \log \mathdotM \pm 0.70,~~{\rm for~\tau_{d}|_{<0.15D}~AGNs}
\end{empheq}
\label{eqn_reg1}
\end{subequations}
with intrinsic scatters of $\epsilon=(0.37, 0.45)$ for (\ref{eqn_reg1}{\it a}, \ref{eqn_reg1}{\it b}), respectively. 
The results of the linear regression for $\Sigma_{\rm d}$ are 
\begin{subequations}
\begin{empheq}[left={\log \Sigma_{\rm d}=\empheqlbrace}]{align}
&(-2.51 \pm 0.31) + (0.20 \pm 0.04) \log \bhm~~({\rm for}~\tau_{\rm d}|_{<0.1D}), \\
%&(-5.18 \pm 1.69) + (0.09 \pm 0.04) \log L_{5100} \pm 0.24,~~{\rm for~\tau_{d}|_{<0.15D}~AGNs}\\
&(-1.08 \pm 0.04) - (0.13 \pm 0.03) \log \mathdotM~~({\rm for}~\tau_{\rm d}|_{<0.1D}), 
\end{empheq}
\label{eqn_reg2}
\end{subequations}
with intrinsic scatters of $\epsilon=(0.18, 0.20)$ for (\ref{eqn_reg2}{\it a}, \ref{eqn_reg2}{\it b}), respectively. 
The results of the linear regression for $F_{\rm var}$ are
\begin{subequations}
\begin{empheq}[left={\log F_{\rm var}=\empheqlbrace}]{align}
&(-2.45 \pm 0.34) + (0.18 \pm 0.05) \log \bhm~~({\rm for}~\tau_{\rm d}|_{<0.1D}), \\
%&(-3.43 \pm 1.76) + (0.05 \pm 0.04) \log L_{5100} \pm 0.28,~~{\rm for~\tau_{d}|_{<0.15D}~AGNs}\\
&(-1.13 \pm 0.04) - (0.13 \pm 0.03) \log \mathdotM~~({\rm for}~\tau_{\rm d}|_{<0.1D}), \\
&(-2.32 \pm 0.22) + (0.16 \pm 0.03) \log \bhm~~({\rm for~all~AGNs}), \\
%&(-1.11 \pm 1.25) + (0.001 \pm 0.03) \log L_{5100} \pm 0.25,~~{\rm for~all~RM~AGNs}\\
&(-1.08 \pm 0.02) - (0.09 \pm 0.02) \log \mathdotM~~({\rm for~all~AGNs}), 
\end{empheq}
\label{eqn_reg3}
\end{subequations}
with intrinsic scatters of $\epsilon=(0.22, 0.21, 0.22, 0.22)$ for 
(\ref{eqn_reg3}{\it a}, \ref{eqn_reg3}{\it b}, \ref{eqn_reg3}{\it c}, \ref{eqn_reg3}{\it d}), respectively. 
From Equation~(\ref{eqn_reg3}), we found that the linear regression results between variability amplitude 
$F_{\rm var}$ and AGN parameters ($\bhm, \mathdotM$ and $L_{5100}$) 
for $\tau_{\rm d}|_{<0.1D}$ sample and all RM AGNs are consistent within the uncertainties. 
For the $\tau_{\rm d}|_{<0.1D}$ sample, because $\Sigma_{\rm d}$ is one-to-one correspondence with $F_{\rm var}$ 
(see Figure~\ref{fig:valid}), their linear regression results with AGN properties ($\bhm$ and $\mathdotM$) 
are consistent within the uncertainties (Equation~\ref{eqn_reg2} and \ref{eqn_reg3}). 
Because the variability amplitudes ($\Sigma_{\rm d}$ and $F_{\rm var}$) do not show significant correlation with optical luminosity 
(for $\Sigma_{\rm d}$: $\rho=0.08$, $p=0.65$; for $F_{\rm var}$: $\rho=0.11$, $p=0.55$), 
and damped variability timescales ($\tau_{\rm d}$) do not show significant correlation with accretion rates 
($\rho=-0.16$, $p=0.34$, and $\beta=-0.15\pm0.10$, $\epsilon=0.60$), 
we do not include its regression results in Equations~(\ref{eqn_reg1}), (\ref{eqn_reg2}) and (\ref{eqn_reg3}). 

For $\tau_{\rm d}|_{<0.1D}$ AGNs, 
the results of correlation analysis (see Figures~\ref{fig:dp-ap}, ~\ref{fig:var-ap} and Table~\ref{tab:pr}) 
and linear regression (Equations~\ref{eqn_reg1}, \ref{eqn_reg2} and \ref{eqn_reg3}) 
between variability characteristics ($\tau_{\rm d}$, $\Sigma_{\rm d}$ and $F_{\rm var}$) and AGN properties show that: 
\begin{enumerate}

\item
Variability amplitudes ($\Sigma_{\rm d}$ and $F_{\rm var}$) do not significantly correlate with optical luminosity, 
which is consistent with the previous results (e.g., \citealt{Bonoli1979,Giallongo1991,Cimatti1993,Netzer1996}). 
Meanwhile, $\Sigma_{\rm d}$ and $F_{\rm var}$ increase with increasing black hole mass, 
which is similar to the results of \cite{Wold2007}, \cite{Wilhite2008}, and \cite{Bauer2009} 
but different from the findings of \cite{Simm2016} and \cite{Sanchez2018}, 
and they decrease with increasing accretion rates, which is consistent with recent results of 
\cite{Simm2016,Rakshit2017} and \cite{Sanchez2018}. 
The latter relation demonstrates that AGN optical fluctuations depend on  accretion rates 
(also see \citealt{Wold2007,Zuo2012,Rakshit2017}). 

\item 
\cite{MacLeod2010} modeled the variability of SDSS stripe 82 (S82) quasars using the DRW model
and found that variability timescale increases with increasing black hole mass with a slope of $0.21\pm0.07$, 
and is nearly independent on optical luminosity.  
However, in our $\tau_{\rm d}|_{<0.1D}$ sample, we found that damped variability timescale $\tau_{\rm d}$ 
increases with increasing black hole mass with a slope of $0.60\pm0.09$, 
and with increasing optical luminosity with a slope of $0.46\pm0.09$. 
Our results give more significant correlations between damped variability timescale and AGN properties than \cite{MacLeod2010}. 
These results, constructed from the different databases, are not strictly consistent with each other, 
which may be attributed to the following reasons: 
1) The sample size of S82 quasars is much larger than RM AGNs, 
but the light curves of RM AGNs have better sampling than S82 quasars. 
2) The light curves of S82 quasars are the photometric data in the $ugriz$ band, 
which includes the contributions of the broad-emission line to some degree. 
Whereas, the light curves of RM AGNs were measured from the optical spectra at 5100~\AA, 
which are not contaminated by broad-emission lines (see Section~\ref{sec:sample}). 
3) The physical properties of RM AGNs (see Section~\ref{sec:sample}) could be more precise than S82 quasars. 
For example, virial BH masses of S82 quasars based on single-epoch spectra 
could include the scatter of the empirical relationship 
(e.g., the scatter of the latest Radius$-$Luminosity relation is larger than 0.3~dex, see \citealt{Du2018}). 

\item 
For sub-Eddington or super-Eddington accretion AGNs, we found from the above relationships that the variability characteristics are generally driven by AGN properties in the same fashion. 
This probably indicates that the variability of AGNs powered by different accretion rates may be caused by the same physical mechanism. 

\end{enumerate}

\begin{figure*}[t!]
\centering
\includegraphics[angle=0,width=1.00\textwidth]{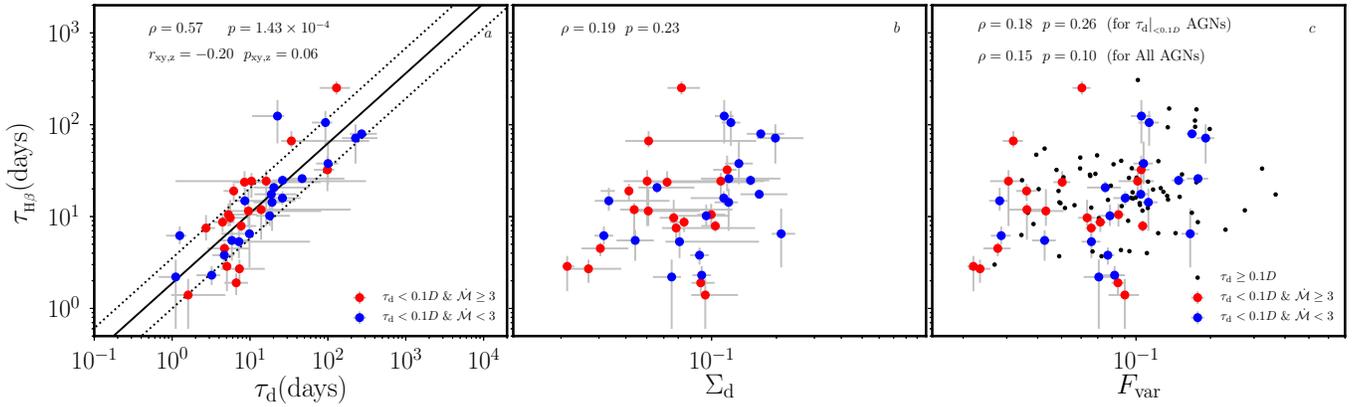}
\caption{\footnotesize 
H$\beta$ time lag ($\tau_{\rm H\beta}$) as a function of damped variability timescale ($\tau_{\rm d}$), 
damped variability amplitude ($\Sigma_{\rm d}$) and model-independent amplitude ($F_{\rm var}$) from the left to right panels. 
In panel ({\it a}), we give the partial correlation coefficient $r_{\rm xy, z}=-0.20$ (where $x \equiv \tau_{\rm d}$, $y \equiv \tau_{\rm H\beta}$ and $z \equiv L_{5100}$) 
and null probability $p_{\rm xy, z}=0.06$. 
In panel ({\it c}), we over-plotted the $\tau_{\rm d}|_{\ge0.1D}$ sample in black dot. 
}
\label{fig:varimla}
\end{figure*}

\subsection{$\rm {H}\beta$ Lag and Variability Characteristics} 
\label{sec:LagV}
In this section, we investigate the relationship between H$\beta$ time lag (i.e., radius of the BLR) 
and variability characteristics (see Figure~\ref{fig:varimla}). 
We found that H$\beta$ time lags do not significantly correlate with the variability amplitudes ($\Sigma_{\rm d}$ and $F_{\rm var}$), 
but significantly correlate with damped variability timescales ($\rho=0.57$, $p=1.43\times10^{-4}$). 
By a linear regression analysis, we yield 
\begin{eqnarray}
\log \tau_{\rm H\beta} = (0.27 \pm 0.13) + (0.76 \pm 0.10) \log \tau_{_{\rm d}}~~({\rm for}~\tau_{\rm d}|_{<0.1D}). \ %\ {\rm [day]}
\label{eqn_reg7}
\end{eqnarray}
This relationship gives an intrinsic scatter of $\epsilon=0.28$~dex 
relative to the dynamic range $\sim$100 of $\tau_{\rm d}$ and $\tau_{\rm H\beta}$. 
In panel ({\it a}) of Figure~\ref{fig:varimla}, the solid line is the best fit and dotted lines are the intrinsic scatter.  
This test shows that the damped variability timescales positively correlate with 
H$\beta$ lags with the slope 0.76 ($r=0.57$, Null probability of $p=1.43 \times 10^{-4}$). 
Another interesting result is that the damped variability timescales are approximately 
equal to H$\beta$ lags ($\tau_{\rm d}/\tau_{\rm H\beta}=1.14\pm0.32$). 
\cite{Kelly2009} suggested that the variability timescales are consistent 
with orbital timescales or thermal timescales of accretion disk. In this case, 
the variability timescales may correspond to the accretion disk scale. 
Therefore, it is possible that $\tau_{\rm d}-\tau_{\rm H\beta}$ relation provides a new insight 
to investigate the connections between the accretion disk and BLR. 
Such as, the BLR size increasing with increasing accretion disk scale. 

On the other hand, the variable optical continuum (originating from the accretion disk) is potentially contaminated by 
nondisk optical continuum emitted from the dense BLR clouds (e.g., see \citealt{Baskin2014,Korista2001}). 
This physical scenario is confirmed by recent work developed by \cite{Chelouche2019}, 
who performed an RM campaign of accretion disk based on the single-source of Mrk~279, 
argued that time delays between adjacent optical bands are associated with the reprocessing of light by nondisk component, 
and suggested that the optical phenomenology of some AGNs may be substantially affected by nondisk continuum emission. 
If this scenario holds in our sample, the above findings may provide another possibility that 
the optical continuum may include the continuum emission from the dense BLR clouds. 
While a fully appreciated nondisk component (i.e. the dense BLR model), which 
emits nondisk continuum, is need to quantitatively investigate this possibility. 

In addition, considering the fact that the damped variability timescale and H$\beta$ time lag positively correlate 
with optical luminosity (see Figure~\ref{fig:dp-ap}{\it a} and $R_{\rm H\beta}-L_{5100}$ relationship of \citealt{Du2018}), 
we employed a partial correlation analysis to investigate the relation of $\tau_{\rm d}-\tau_{\rm H\beta}-L_{5100}$. 
The correlation coefficient 
between $\mathbi{x}$ and $\mathbi{y}$ excluding the dependence on the third parameter of $\mathbi{z}$ 
is evaluated as (e.g., \citealt{Kendall1979}) 
\begin{equation}
 {r_{xy,z}}=\frac{{r_{xy}}-{r_{xz}r_{yz}}}{\sqrt{1-{r_{xz}}^{2}}\sqrt{1-{r_{yz}}^2}}, 
\end{equation}
where ${r_{xy},~r_{xz}~{\rm and}~r_{yz}}$ is the Spearman rank-order correlation coefficient 
between $\mathbi{x}$ and $\mathbi{y}$, between $\mathbi{x}$ and $\mathbi{z}$ and between $\mathbi{y}$ 
and $\mathbi{z}$ \citep{Press1992}, respectively. 
The partial correlation coefficient between $\tau_{\rm H\beta}$ and $\tau_{\rm d}$ excluding the dependence on $L_{5100}$ 
is $r_{\rm xy, z}=-0.20$ (where $x \equiv \tau_{\rm d}$, $y \equiv \tau_{\rm H\beta}$ and $z \equiv L_{5100}$) 
with a null probability of $p_{\rm xy, z}=0.06$.  
In some degree, the optical luminosity could modulate the variability timescale and the BLR size in the same pattern. 
Anyways, $\tau_{\rm d}$ and $\tau_{\rm H\beta}$ relationship with reasonable scatter (0.28~dex) provides a possible 
to estimate the BLR size using a large amount of time-domain data of AGNs in the near future. 

\section{Conclusion}
\label{sec:con}
We employed the DRW model and the traditional method to quantify the optical variability 
characteristics of all RM AGNs. 
The DRW model is described by damped variability timescale and variability amplitude. 
The traditional method gives model-independent variability amplitude $R_{\rm max}$ and $F_{\rm var}$.  
We checked the validity of the damped model parameters 
comparing damped variability amplitude with intrinsic variability amplitude, 
and found that damped variability amplitude 
only for the AGNs with $\tau_{\rm d}$ less than 10\% of observation length (i.e. $\tau_{\rm d}|_{<0.1D}$) 
is linearly correlated with the intrinsic variability amplitude 
(slope $\beta=1.05\pm0.06$, intrinsic scatter $\epsilon=0.03$, 
and correlation coefficients approximately equal to 1). 
Meanwhile, we found that only 35\% of light curves can return reliable 
DRW model parameters (Section~\ref{sec:valid}) in RM AGNs, 
this percentage is significantly lower than \cite{Sanchez2018} suggestion (74\%) based on their selected sample. 
Therefor, we only adopted $\tau_{\rm d}|_{<0.1D}$ AGNs to investigate variability characteristics. 
Our main results are summarised as follows: 

\begin{itemize}
%\begin{enumerate}
\item 
Employing the Kolmogorov-Smirnov statistic test, we found that the model-independent variability characteristics 
($F_{\rm var}$ and $R_{\rm max}$ of all RM AGNs) 
of $\mathdotM<3$ and $\mathdotM\geq3$ AGNs have significantly different distributions, 
and the model-dependent variability characteristics ($\Sigma_{\rm d},\tau_{\rm d}$) 
of $\mathdotM<3$ and $\mathdotM\geq3$ AGNs have marginally different distributions.  

\item
We found that the variability characteristics of the sub-Eddington and super-Eddington accretion AGNs 
are generally correlated with AGN parameters in the same fashion, 
which might indicate that the variability of AGNs powered by different accretion rates 
may be caused by the same physical mechanism. 
To be specific, the damped variability timescales are positively correlated 
with BH mass and optical luminosity, but weakly anticorrelated with accretion rates. 
The variability amplitudes are positively correlated with BH mass 
and anticorrelated with accretion rates, but not correlated with optical luminosity. 

\item
We found that H$\beta$ lags ($\tau_{\rm H\beta}$) are not correlated with the variability amplitudes, 
but positively correlated with damped variability timescales ($\tau_{\rm d}$) with a scatter of 0.28~dex (see Section~\ref{sec:LagV}). 
Meanwhile, we also found that the partial  correlation coefficient between $\tau_{\rm H\beta}$ and $\tau_{\rm d}$ 
excluding the dependence on optical luminosity $L_{5100}$ is $-0.20$. 
We found $\tau_{\rm d}=(1.14\pm0.32)\tau_{\rm H\beta}$, 
which could indicate that the BLR size is correlated with accretion disk scale 
if the variability timescales correspond to the accretion disk scale. 
\end{itemize}

\acknowledgements
{We are grateful to the referee for constructive suggestions that significantly improved the manuscript. 
We acknowledge the support of the staff of the Lijiang 2.4m telescope. 
Funding for the telescope has been provided by CAS and the People's Government of Yunnan Province. 
This research is supported in part by National Key Program for Science and Technology Research and Development of China 
(grants 2016YFA0400701 and 2016YFA0400702), by grants NSFC-11833008, -11173023, -11133006, -11373024, -11473002, 
and by Key Research Program of Frontier Sciences, CAS, Grant QYZDJ-SSW-SLH007. 
K.X.L. acknowledges financial support from the National Natural Science Foundation of China (No. NSFC-11703077) 
and from the Light of West China Program provided by CAS (No. Y7XB016001). 
}

%bibliography
%\input{bib.bbl}
%

\begin{deluxetable*}{lcccccccccc}
\tablecolumns{11}
\tablewidth{0pt}
\setlength{\tabcolsep}{4pt}
\tablecaption{RM AGN Properties and Variability Characteristics \label{table:drw}}
\tabletypesize{\scriptsize}
\tablehead{
\colhead{Object}            &
\colhead{$z$}               &
\colhead{$\log (L_{5100} / \ergs)$}    &
\colhead{$\log (\bhm / \msun)$} &
\colhead{$\log \mathdotM$}  &
\colhead{$F_{\rm var}(\%)$}         &
\colhead{$R_{\rm max}$}         &
\colhead{$\log \Sigma_{\rm d}$} &
\colhead{$\log$ ( $\tau_{\rm d} /$days)}   &
\colhead{$\tau_{\rm d}/D$} &
\colhead{Ref.}                  \\
\colhead{(1)}&
\colhead{(2)}&
\colhead{(3)}&
\colhead{(4)}&
\colhead{(5)}&
\colhead{(6)}&
\colhead{(7)}&
\colhead{(8)}&
\colhead{(9)}&
\colhead{(10)}&
\colhead{(11)}
}
\startdata
\multicolumn{10}{c}{SEAMBH2012}\\ \cline{1-11}  
   Mrk       335&  0.0258&$ 43.69\pm  0.06$&$  6.87_{-  0.14}^{+  0.10}$&$  1.17_{-  0.30}^{+  0.31}$&$  7.17\pm  0.38$&  1.48&$ -1.13^{+  0.02}_{-  0.07}$&$  0.65^{+  0.15}_{-  0.11}$&  0.04 &19,21\\%                                    lj12
   Mrk      1044&  0.0165&$ 43.10\pm  0.10$&$  6.45_{-  0.13}^{+  0.12}$&$  1.22_{-  0.41}^{+  0.40}$&$  8.49\pm  0.49$&  1.81&$ -1.00^{+  0.07}_{-  0.07}$&$  0.72^{+  0.37}_{-  0.11}$&  0.05 &19,21\\%                                    lj12
  IRAS     04416+1215&  0.0889&$ 44.47\pm  0.03$&$  6.78_{-  0.06}^{+  0.31}$&$  2.63_{-  0.67}^{+  0.16}$&$  5.64\pm  0.32$&  1.31&$ -1.15^{+  0.41}_{-  0.07}$&$  1.21^{+  0.98}_{-  0.15}$&  0.11 &19,21\\%                                    lj12
   Mrk       382&  0.0337&$ 43.12\pm  0.08$&$  6.50_{-  0.29}^{+  0.19}$&$  1.18_{-  0.53}^{+  0.69}$&$  6.59\pm  0.34$&  1.43&$ -1.16^{+  0.37}_{-  0.07}$&$  0.44^{+  0.12}_{-  0.07}$&  0.01 &19,21\\%                                    lj12
   Mrk       142&  0.0449&$ 43.56\pm  0.06$&$  6.59_{-  0.07}^{+  0.07}$&$  1.90_{-  0.86}^{+  0.85}$&$ 10.64\pm  0.51$&  1.83&$ -0.98^{+  0.02}_{-  0.07}$&$  0.89^{+  0.20}_{-  0.07}$&  0.05 &19,21\\%                                    lj12
   MCG  +06-26-012&  0.0328&$ 42.67\pm  0.11$&$  6.92_{-  0.12}^{+  0.14}$&$ -0.34_{-  0.45}^{+  0.37}$&$ 11.18\pm  1.04$&  2.02&$ -0.76^{+  0.63}_{-  0.11}$&$  1.43^{+  1.17}_{-  0.22}$&  0.19 &19,21\\%                                    lj12
  IRAS    F12397+3333&  0.0435&$ 44.23\pm  0.05$&$  6.79_{-  0.45}^{+  0.27}$&$  2.26_{-  0.62}^{+  0.98}$&$  6.34\pm  0.51$&  1.29&$ -1.18^{+  0.02}_{-  0.07}$&$  0.75^{+  0.24}_{-  0.11}$&  0.04 &19,21\\%                                    lj12
   Mrk       486&  0.0389&$ 43.69\pm  0.05$&$  7.24_{-  0.06}^{+  0.12}$&$  0.55_{-  0.32}^{+  0.20}$&$  5.03\pm  0.40$&  1.24&$ -1.21^{+  0.41}_{-  0.11}$&$  0.93^{+  0.98}_{-  0.15}$&  0.08 &19,21\\%                                    lj12
   Mrk       493&  0.0313&$ 43.11\pm  0.08$&$  6.14_{-  0.11}^{+  0.04}$&$  1.88_{-  0.21}^{+  0.33}$&$ 10.62\pm  1.03$&  1.77&$ -0.80^{+  0.76}_{-  0.11}$&$  1.29^{+  1.54}_{-  0.46}$&  0.36 &19,21\\%                                    lj12
   \hline
   \multicolumn{10}{c}{SEAMBH2013}\\ \cline{1-11}  
  SDSS   J075101&  0.1208&$ 44.12\pm  0.05$&$  7.16_{-  0.09}^{+  0.17}$&$  1.34_{-  0.41}^{+  0.25}$&$  5.94\pm  0.69$&  1.37&$ -0.88^{+  0.50}_{-  0.24}$&$  2.07^{+  1.32}_{-  0.37}$&  0.69 &22\\%                                    lj13
  SDSS   J080101&  0.1398&$ 44.27\pm  0.03$&$  6.78_{-  0.17}^{+  0.34}$&$  2.33_{-  0.72}^{+  0.39}$&$  3.85\pm  0.42$&  1.15&$ -1.38^{+  0.67}_{-  0.11}$&$  1.51^{+  1.45}_{-  0.24}$&  0.23 &22\\%                                    lj13
  SDSS   J080131&  0.1786&$ 43.98\pm  0.04$&$  6.50_{-  0.16}^{+  0.24}$&$  2.46_{-  0.54}^{+  0.38}$&$  4.32\pm  0.78$&  1.36&$ -1.29^{+  0.24}_{-  0.07}$&$  0.98^{+  0.93}_{-  0.28}$&  0.07 &22\\%                                    lj13
  SDSS   J081441&  0.1628&$ 44.01\pm  0.07$&$  6.97_{-  0.27}^{+  0.23}$&$  1.56_{-  0.57}^{+  0.63}$&$  9.68\pm  0.94$&  1.47&$ -0.83^{+  0.54}_{-  0.15}$&$  1.98^{+  1.15}_{-  0.37}$&  0.56 &22\\%                                    lj13
  SDSS   J081456&  0.1197&$ 43.99\pm  0.04$&$  7.44_{-  0.49}^{+  0.12}$&$  0.59_{-  0.30}^{+  1.03}$&$  3.05\pm  0.48$&  1.19&$ -1.30^{+  0.54}_{-  0.11}$&$  1.21^{+  1.28}_{-  0.20}$&  0.10 &22\\%                                    lj13
  SDSS   J093922&  0.1862&$ 44.07\pm  0.04$&$  6.53_{-  0.33}^{+  0.07}$&$  2.54_{-  0.20}^{+  0.71}$&$  3.62\pm  0.56$&  1.23&$ -1.36^{+  0.46}_{-  0.07}$&$  1.14^{+  1.15}_{-  0.20}$&  0.08 &22\\%                                    lj13
  \hline
  \multicolumn{10}{c}{SEAMBH2014}\\ \cline{1-11}  
  SDSS   J075949&  0.1879&$ 44.20\pm  0.03$&$  7.54_{-  0.12}^{+  0.12}$&$  0.70_{-  0.29}^{+  0.28}$&$  4.27\pm  0.39$&  1.22&$ -1.14^{+  0.59}_{-  0.15}$&$  1.91^{+  1.28}_{-  0.33}$&  0.51 &23\\%                                    lj14
  SDSS   J080131&  0.1784&$ 43.95\pm  0.04$&$  6.56_{-  0.90}^{+  0.37}$&$  2.29_{-  0.80}^{+  1.87}$&$  4.26\pm  0.45$&  1.20&$ -1.33^{+  0.63}_{-  0.11}$&$  1.75^{+  1.32}_{-  0.33}$&  0.44 &23\\%                                    lj14
  SDSS   J084533&  0.3024&$ 44.54\pm  0.03$&$  6.66_{-  0.23}^{+  0.08}$&$  2.98_{-  0.23}^{+  0.51}$&$  7.70\pm  0.68$&  1.32&$ -0.87^{+  0.54}_{-  0.24}$&$  2.12^{+  1.17}_{-  0.39}$&  1.13 &23\\%                                    lj14
  SDSS   J085946&  0.2440&$ 44.41\pm  0.03$&$  6.80_{-  0.12}^{+  0.25}$&$  2.51_{-  0.53}^{+  0.29}$&$  3.65\pm  0.43$&  1.21&$ -1.36^{+  0.59}_{-  0.11}$&$  1.50^{+  1.28}_{-  0.20}$&  0.21 &23\\%                                    lj14
  SDSS   J102339&  0.1364&$ 44.09\pm  0.03$&$  7.16_{-  0.07}^{+  0.26}$&$  1.29_{-  0.55}^{+  0.20}$&$  3.46\pm  0.32$&  1.18&$ -1.33^{+  0.59}_{-  0.11}$&$  1.75^{+  1.24}_{-  0.28}$&  0.33 &23\\%                                    lj14
  \hline
  \multicolumn{10}{c}{SEAMBH2015-2016}\\ \cline{1-11}  
  SDSS   J074352&  0.2520&$ 45.37\pm  0.02$&$  7.93_{-  0.04}^{+  0.05}$&$  1.69_{-  0.13}^{+  0.12}$&$  5.95\pm  0.50$&  1.23&$ -1.03^{+  0.41}_{-  0.20}$&$  2.26^{+  0.83}_{-  0.22}$&  1.01 &24\\%                                    lj16
  SDSS   J075051&  0.4004&$ 45.33\pm  0.01$&$  7.67_{-  0.07}^{+  0.11}$&$  2.14_{-  0.24}^{+  0.16}$&$  3.19\pm  0.30$&  1.18&$ -1.29^{+  0.50}_{-  0.11}$&$  1.53^{+  1.00}_{-  0.13}$&  0.08 &24\\%                                    lj56
  SDSS   J075101&  0.1209&$ 44.24\pm  0.04$&$  7.20_{-  0.12}^{+  0.08}$&$  1.45_{-  0.23}^{+  0.30}$&$  6.51\pm  0.82$&  1.27&$ -0.95^{+  0.46}_{-  0.15}$&$  2.40^{+  0.91}_{-  0.30}$&  1.26 &24\\%                                    lj16
  SDSS   J075949&  0.1879&$ 44.19\pm  0.06$&$  7.21_{-  0.19}^{+  0.16}$&$  1.34_{-  0.42}^{+  0.48}$&$  9.36\pm  1.12$&  1.38&$ -1.03^{+  0.50}_{-  0.11}$&$  1.78^{+  1.00}_{-  0.13}$&  0.14 &24\\%                                    lj56
  SDSS   J081441&  0.1626&$ 43.95\pm  0.04$&$  7.22_{-  0.11}^{+  0.10}$&$  0.97_{-  0.28}^{+  0.28}$&$  5.09\pm  0.76$&  1.21&$ -1.19^{+  0.67}_{-  0.11}$&$  1.81^{+  1.26}_{-  0.30}$&  0.45 &24\\%                                    lj16
  SDSS   J083553&  0.2051&$ 44.44\pm  0.02$&$  6.87_{-  0.25}^{+  0.16}$&$  2.41_{-  0.35}^{+  0.53}$&$  5.16\pm  0.51$&  1.31&$ -0.93^{+  0.50}_{-  0.20}$&$  2.18^{+  1.09}_{-  0.30}$&  0.98 &24\\%                                    lj16
  SDSS   J084533&  0.3024&$ 44.52\pm  0.02$&$  6.82_{-  0.10}^{+  0.14}$&$  2.64_{-  0.31}^{+  0.22}$&$  4.04\pm  0.59$&  1.18&$ -1.29^{+  0.63}_{-  0.11}$&$  1.37^{+  1.26}_{-  0.22}$&  0.18 &24\\%                                    lj16
  SDSS   J093302&  0.1772&$ 44.31\pm  0.13$&$  7.08_{-  0.11}^{+  0.08}$&$  1.79_{-  0.40}^{+  0.40}$&$  3.61\pm  0.34$&  1.18&$ -1.38^{+ -0.13}_{-  0.07}$&$  0.79^{+  0.04}_{-  0.13}$&  0.03 &24\\%                                    lj16
  SDSS   J100402&  0.3272&$ 45.52\pm  0.01$&$  7.44_{-  0.06}^{+  0.37}$&$  2.89_{-  0.75}^{+  0.13}$&$  4.77\pm  0.36$&  1.22&$ -1.21^{+  0.46}_{-  0.11}$&$  2.06^{+  0.91}_{-  0.13}$&  0.69 &24\\%                                    lj16
  SDSS   J101000&  0.2564&$ 44.76\pm  0.02$&$  7.46_{-  0.14}^{+  0.27}$&$  1.70_{-  0.56}^{+  0.31}$&$  6.53\pm  0.56$&  1.30&$ -1.11^{+  0.50}_{-  0.11}$&$  1.84^{+  1.00}_{-  0.13}$&  0.41 &24\\%                                    lj16
   \hline
    3C       120&  0.0330&$ 44.07\pm  0.05$&$  7.61_{-  0.22}^{+  0.19}$&$  0.03_{-  0.37}^{+  0.37}$&$ 10.46\pm  0.56$&  1.49&$ -0.60^{+  0.12}_{-  0.17}$&$  2.35^{+  0.27}_{-  0.21}$&  1.66 &1\\%                 3c120_con_grier2012.txt
    3C       120&  0.0330&$ 43.94\pm  0.05$&$  7.80_{-  0.06}^{+  0.05}$&$ -0.17_{-  0.37}^{+  0.37}$&$ 17.81\pm  1.78$&  2.34&$ -0.92^{+  0.24}_{-  0.13}$&$  1.67^{+  0.54}_{-  0.30}$&  0.02 &2\\%                  3c120_peterson1998.txt
    3C     390.3&  0.0561&$ 43.68\pm  0.10$&$  8.87_{-  0.15}^{+  0.10}$&$ -3.35_{-  0.65}^{+  0.60}$&$ 17.54\pm  1.55$&  2.63&$ -0.21^{+  0.17}_{-  0.30}$&$  3.23^{+  0.43}_{-  0.57}$&  3.23 &3\\%              3c390_con_dietrich1998.txt
    3C     390.3&  0.0561&$ 44.50\pm  0.03$&$  9.20_{-  0.03}^{+  0.03}$&$ -2.12_{-  0.51}^{+  0.51}$&$  6.96\pm  0.95$&  1.29&$ -0.80^{+  0.24}_{-  0.26}$&$  2.38^{+  0.77}_{-  0.45}$&  3.03 &16\\%              3c390_con_dietrich2012.txt
   Arp       151&  0.0211&$ 42.55\pm  0.10$&$  6.87_{-  0.08}^{+  0.05}$&$ -0.44_{-  0.28}^{+  0.30}$&$ 11.06\pm  0.91$&  1.54&$ -0.70^{+  0.15}_{-  0.22}$&$  1.87^{+  0.46}_{-  0.27}$&  0.79 &4\\%                arp151_con_bentz2009.txt
Fairal         9&  0.0470&$ 43.98\pm  0.04$&$  8.09_{-  0.12}^{+  0.07}$&$ -0.71_{-  0.21}^{+  0.31}$&$ 36.68\pm  3.26$&  3.71&$ -0.30^{+  0.34}_{-  0.22}$&$  2.50^{+  0.46}_{-  0.50}$&  1.38 &5\\%             fairall9_con_santos1997.txt
   Mrk      1310&  0.0196&$ 42.29\pm  0.14$&$  6.62_{-  0.08}^{+  0.07}$&$ -0.31_{-  0.39}^{+  0.35}$&$  7.25\pm  0.72$&  1.39&$ -1.02^{+  0.33}_{-  0.09}$&$  1.30^{+  0.72}_{-  0.30}$&  0.20 &4\\%               mrk1310_con_bentz2009.txt
   Mrk       142&  0.0449&$ 43.61\pm  0.04$&$  6.06_{-  0.16}^{+  0.10}$&$  1.96_{-  0.82}^{+  0.82}$&$  2.34\pm  0.24$&  1.12&$ -1.57^{+  0.15}_{-  0.09}$&$  0.86^{+  0.33}_{-  0.24}$&  0.07 &4\\%                mrk142_con_bentz2009.txt
   Mrk       202&  0.0210&$ 42.26\pm  0.14$&$  6.11_{-  0.20}^{+  0.20}$&$  0.66_{-  0.65}^{+  0.59}$&$  2.68\pm  0.33$&  1.18&$ -1.50^{+  0.12}_{-  0.16}$&$  1.03^{+  0.52}_{-  0.28}$&  0.10 &4\\%                mrk202_con_bentz2009.txt
   Mrk       290&  0.0296&$ 43.17\pm  0.06$&$  7.55_{-  0.07}^{+  0.07}$&$ -0.85_{-  0.23}^{+  0.23}$&$ 17.78\pm  0.98$&  2.18&$ -0.60^{+  0.28}_{-  0.20}$&$  2.11^{+  0.54}_{-  0.37}$&  0.94 &6\\%               mrk290_con_denney2010.txt
   Mrk       509&  0.0344&$ 44.19\pm  0.05$&$  8.15_{-  0.03}^{+  0.03}$&$ -0.52_{-  0.14}^{+  0.13}$&$ 16.84\pm  0.88$&  2.15&$ -0.77^{+  0.11}_{-  0.07}$&$  2.43^{+  0.20}_{-  0.15}$&  0.09 &2\\%                 mrk509_peterson1998.txt
   NGC      3227&  0.0039&$ 42.24\pm  0.11$&$  7.09_{-  0.12}^{+  0.09}$&$ -1.34_{-  0.36}^{+  0.38}$&$  7.69\pm  0.47$&  1.60&$ -1.06^{+  0.04}_{-  0.04}$&$  0.67^{+  0.13}_{-  0.15}$&  0.05 &6\\%              ngc3227_con_denney2010.txt
   NGC      3516&  0.0088&$ 42.79\pm  0.20$&$  7.82_{-  0.08}^{+  0.05}$&$ -1.97_{-  0.52}^{+  0.41}$&$ 27.53\pm  1.44$&  5.90&$ -0.32^{+  0.22}_{-  0.20}$&$  2.20^{+  0.24}_{-  0.21}$&  1.33 &6\\%              ngc3516_con_denney2010.txt
   NGC      3783&  0.0097&$ 42.56\pm  0.18$&$  7.45_{-  0.11}^{+  0.12}$&$ -1.58_{-  0.59}^{+  0.45}$&$  7.84\pm  0.81$&  1.52&$ -1.02^{+  0.11}_{-  0.13}$&$  1.25^{+  0.26}_{-  0.17}$&  0.08 &7\\%              ngc3783_con_stirpe1994.txt
   NGC      4151&  0.0033&$ 42.09\pm  0.21$&$  7.72_{-  0.06}^{+  0.07}$&$ -2.81_{-  0.57}^{+  0.37}$&$ 12.29\pm  1.38$&  1.47&$ -0.74^{+  0.17}_{-  0.20}$&$  1.66^{+  0.37}_{-  0.33}$&  1.11 &8\\%               ngc4151_con_bentz2006.txt
   NGC      4253&  0.0129&$ 42.57\pm  0.12$&$  6.49_{-  0.10}^{+  0.10}$&$  0.36_{-  0.42}^{+  0.36}$&$  2.85\pm  0.26$&  1.15&$ -1.50^{+  0.04}_{-  0.04}$&$  0.10^{+  0.09}_{-  0.17}$&  0.01 &4\\%               ngc4253_con_bentz2009.txt
   NGC      4593&  0.0090&$ 42.87\pm  0.18$&$  7.28_{-  0.10}^{+  0.08}$&$ -0.73_{-  0.52}^{+  0.41}$&$  4.89\pm  0.72$&  1.25&$ -1.21^{+  0.33}_{-  0.13}$&$  1.31^{+  0.63}_{-  0.41}$&  0.50 &9\\%              ngc4593_con_denney2006.txt
   NGC      4748&  0.0146&$ 42.56\pm  0.12$&$  6.61_{-  0.23}^{+  0.11}$&$  0.10_{-  0.44}^{+  0.61}$&$  4.27\pm  0.47$&  1.18&$ -1.35^{+  0.09}_{-  0.09}$&$  0.77^{+  0.22}_{-  0.17}$&  0.05 &4\\%               ngc4748_con_bentz2009.txt
   NGC      6814&  0.0052&$ 42.12\pm  0.28$&$  7.16_{-  0.06}^{+  0.05}$&$ -1.64_{-  0.80}^{+  0.46}$&$ 14.45\pm  1.53$&  1.68&$ -0.75^{+  0.28}_{-  0.14}$&$  1.31^{+  0.61}_{-  0.24}$&  0.28 &4\\%               ngc6814_con_bentz2009.txt
   NGC      7469&  0.0163&$ 43.32\pm  0.12$&$  6.42_{-  0.09}^{+  0.06}$&$  0.63_{-  1.90}^{+  1.85}$&$  2.76\pm  0.35$&  1.10&$ -1.52^{+  0.13}_{-  0.09}$&$  0.67^{+  0.33}_{-  0.22}$&  0.08 &10\\%             ngc7469_con_collier1998.txt
    PG  2130+099&  0.0630&$ 44.20\pm  0.03$&$  7.05_{-  0.10}^{+  0.08}$&$  1.69_{-  0.20}^{+  0.23}$&$  7.07\pm  0.34$&  1.33&$ -0.78^{+  0.24}_{-  0.24}$&$  2.57^{+  0.46}_{-  0.52}$&  3.08 &1\\%                pg2130_con_grier2012.txt
   SBS 1116+583A&  0.0279&$ 42.14\pm  0.23$&$  6.78_{-  0.12}^{+  0.11}$&$ -0.87_{-  0.71}^{+  0.51}$&$  8.19\pm  0.83$&  1.47&$ -1.05^{+  0.04}_{-  0.04}$&$  0.51^{+  0.11}_{-  0.20}$&  0.03 &4\\%               sbs1116_con_bentz2009.txt
   NGC      5548&  0.0172&$ 42.96\pm  0.13$&$  7.70_{-  0.20}^{+  0.15}$&$ -2.27_{-  0.55}^{+  0.49}$&$  3.44\pm  0.86$&  1.39&$ -1.39^{+  0.37}_{-  0.13}$&$  1.31^{+  0.67}_{-  0.56}$&  0.52 &11\\%               ngc5548_con_bentz2007.txt
   NGC      5548&  0.0172&$ 43.01\pm  0.11$&$  8.12_{-  0.16}^{+  0.08}$&$ -2.19_{-  0.51}^{+  0.47}$&$  9.40\pm  0.90$&  1.40&$ -0.87^{+  0.30}_{-  0.13}$&$  1.55^{+  0.61}_{-  0.33}$&  0.33 &4\\%               ngc5548_con_bentz2009.txt
   NGC      5548&  0.0172&$ 42.99\pm  0.11$&$  8.50_{-  0.17}^{+  0.09}$&$ -2.21_{-  0.53}^{+  0.48}$&$ 11.16\pm  0.60$&  1.71&$ -0.68^{+  0.28}_{-  0.20}$&$  1.85^{+  0.61}_{-  0.33}$&  0.48 &6\\%              ngc5548_con_denney2010.txt
   NGC      5548&  0.0172&$ 43.21\pm  0.12$&$  7.94_{-  0.13}^{+  0.16}$&$ -1.68_{-  0.50}^{+  0.48}$&$ 23.00\pm  2.00$&  2.31&$ -0.61^{+  0.46}_{-  0.11}$&$  1.48^{+  0.71}_{-  0.11}$&  0.15 &18\\%                                      lj
   NGC      5548&  0.0172&$ 43.59\pm  0.09$&$  8.30_{-  0.04}^{+  0.07}$&$ -1.33_{-  0.48}^{+  0.45}$&$ 10.05\pm  0.62$&  1.59&$ -0.79^{+  0.15}_{-  0.13}$&$  1.92^{+  0.39}_{-  0.26}$&  0.27 &12\\%          ngc5548_con_peterson2002_10.tx
   NGC      5548&  0.0172&$ 43.51\pm  0.09$&$  8.28_{-  0.06}^{+  0.05}$&$ -1.45_{-  0.48}^{+  0.45}$&$ 14.86\pm  0.92$&  1.98&$ -0.82^{+  0.09}_{-  0.09}$&$  1.42^{+  0.22}_{-  0.13}$&  0.08 &12\\%          ngc5548_con_peterson2002_11.tx
   NGC      5548&  0.0172&$ 43.11\pm  0.11$&$  7.69_{-  0.37}^{+  0.27}$&$ -2.04_{-  0.51}^{+  0.47}$&$ 16.56\pm  1.30$&  2.40&$ -0.68^{+  0.07}_{-  0.04}$&$  0.99^{+  0.05}_{-  0.36}$&  0.04 &12\\%          ngc5548_con_peterson2002_12.tx
   NGC      5548&  0.0172&$ 43.11\pm  0.11$&$  8.07_{-  0.31}^{+  0.15}$&$ -2.03_{-  0.51}^{+  0.47}$&$ 11.22\pm  1.02$&  1.68&$ -0.92^{+  0.04}_{-  0.07}$&$  1.28^{+  0.13}_{-  0.24}$&  0.05 &12\\%          ngc5548_con_peterson2002_13.tx
   NGC      5548&  0.0172&$ 43.39\pm  0.10$&$  7.92_{-  0.04}^{+  0.03}$&$ -1.62_{-  0.49}^{+  0.46}$&$ 11.72\pm  0.87$&  2.16&$ -0.79^{+  0.15}_{-  0.18}$&$  1.84^{+  0.39}_{-  0.24}$&  0.24 &12\\%          ngc5548_con_peterson2002_1.txt
   NGC      5548&  0.0172&$ 43.14\pm  0.11$&$  8.03_{-  0.06}^{+  0.05}$&$ -1.99_{-  0.51}^{+  0.47}$&$ 12.90\pm  1.08$&  1.82&$ -0.49^{+  0.12}_{-  0.25}$&$  2.58^{+  0.39}_{-  0.46}$&  1.23 &12\\%          ngc5548_con_peterson2002_2.txt
   NGC      5548&  0.0172&$ 43.35\pm  0.09$&$  7.93_{-  0.08}^{+  0.07}$&$ -1.68_{-  0.49}^{+  0.45}$&$  9.04\pm  0.94$&  1.51&$ -0.94^{+  0.02}_{-  0.15}$&$  1.42^{+  0.24}_{-  0.20}$&  0.09 &12\\%          ngc5548_con_peterson2002_3.txt
   NGC      5548&  0.0172&$ 43.07\pm  0.11$&$  7.89_{-  0.09}^{+  0.07}$&$ -2.10_{-  0.52}^{+  0.47}$&$ 16.82\pm  1.39$&  2.04&$ -0.42^{+  0.12}_{-  0.25}$&$  2.59^{+  0.33}_{-  0.49}$&  1.45 &12\\%          ngc5548_con_peterson2002_4.txt
   NGC      5548&  0.0172&$ 43.32\pm  0.10$&$  7.95_{-  0.05}^{+  0.05}$&$ -1.72_{-  0.49}^{+  0.46}$&$  8.75\pm  0.53$&  1.65&$ -0.87^{+  0.20}_{-  0.13}$&$  1.95^{+  0.41}_{-  0.26}$&  0.30 &12\\%          ngc5548_con_peterson2002_5.txt
   NGC      5548&  0.0172&$ 43.38\pm  0.09$&$  8.15_{-  0.17}^{+  0.11}$&$ -1.64_{-  0.49}^{+  0.45}$&$ 10.37\pm  0.75$&  1.76&$ -0.78^{+  0.20}_{-  0.13}$&$  1.95^{+  0.43}_{-  0.24}$&  0.28 &12\\%          ngc5548_con_peterson2002_6.txt
   NGC      5548&  0.0172&$ 43.52\pm  0.09$&$  8.31_{-  0.06}^{+  0.05}$&$ -1.43_{-  0.48}^{+  0.45}$&$  7.91\pm  0.68$&  1.48&$ -0.87^{+  0.22}_{-  0.15}$&$  2.07^{+  0.46}_{-  0.30}$&  0.37 &12\\%          ngc5548_con_peterson2002_7.txt
   NGC      5548&  0.0172&$ 43.43\pm  0.09$&$  8.15_{-  0.03}^{+  0.03}$&$ -1.56_{-  0.48}^{+  0.45}$&$ 15.01\pm  0.94$&  1.86&$ -0.53^{+  0.17}_{-  0.22}$&$  2.49^{+  0.37}_{-  0.41}$&  0.95 &12\\%          ngc5548_con_peterson2002_8.txt
   NGC      5548&  0.0172&$ 43.24\pm  0.10$&$  8.13_{-  0.04}^{+  0.05}$&$ -1.85_{-  0.49}^{+  0.46}$&$ 10.45\pm  0.75$&  1.84&$ -0.78^{+  0.13}_{-  0.09}$&$  1.27^{+  0.33}_{-  0.22}$&  0.06 &12\\%          ngc5548_con_peterson2002_9.txt
    PG  0026+129&  0.1420&$ 44.97\pm  0.02$&$  8.15_{-  0.13}^{+  0.09}$&$  0.65_{-  0.20}^{+  0.28}$&$ 17.31\pm  1.27$&  2.22&$ -0.65^{+  0.11}_{-  0.09}$&$  2.38^{+  0.24}_{-  0.20}$&  0.10 &13\\%                pg0026_con_kaspi2000.txt
    PG  0052+251&  0.1544&$ 44.81\pm  0.03$&$  8.64_{-  0.14}^{+  0.11}$&$ -0.59_{-  0.25}^{+  0.31}$&$ 19.86\pm  1.45$&  2.70&$ -0.62^{+  0.13}_{-  0.09}$&$  2.36^{+  0.26}_{-  0.17}$&  0.10 &13\\%                pg0052_con_kaspi2000.txt
    PG  0804+761&  0.1000&$ 44.91\pm  0.02$&$  8.43_{-  0.06}^{+  0.05}$&$  0.00_{-  0.13}^{+  0.15}$&$ 17.55\pm  1.20$&  1.94&$ -0.75^{+  0.15}_{-  0.11}$&$  2.66^{+  0.28}_{-  0.26}$&  0.18 &13\\%                pg0804_con_kaspi2000.txt
    PG  0953+414&  0.2341&$ 45.19\pm  0.01$&$  8.44_{-  0.07}^{+  0.06}$&$  0.39_{-  0.14}^{+  0.16}$&$ 13.61\pm  1.14$&  1.66&$ -0.75^{+  0.17}_{-  0.13}$&$  2.86^{+  0.33}_{-  0.37}$&  0.33 &13\\%                pg0953_con_kaspi2000.txt
    PG  1226+023&  0.1583&$ 45.96\pm  0.02$&$  8.87_{-  0.15}^{+  0.09}$&$  0.70_{-  0.20}^{+  0.33}$&$ 10.17\pm  0.91$&  1.56&$ -0.92^{+  0.13}_{-  0.09}$&$  2.49^{+  0.28}_{-  0.17}$&  0.14 &13\\%                pg1226_con_kaspi2000.txt
    PG  1229+204&  0.0630&$ 43.70\pm  0.05$&$  8.03_{-  0.23}^{+  0.24}$&$ -1.03_{-  0.55}^{+  0.52}$&$ 10.73\pm  0.97$&  1.65&$ -0.87^{+  0.07}_{-  0.07}$&$  2.00^{+  0.13}_{-  0.15}$&  0.04 &13\\%                pg1229_con_kaspi2000.txt
    PG  1307+085&  0.1550&$ 44.85\pm  0.02$&$  8.72_{-  0.26}^{+  0.13}$&$ -0.68_{-  0.28}^{+  0.53}$&$ 11.30\pm  1.08$&  1.71&$ -0.91^{+  0.04}_{-  0.04}$&$  1.97^{+  0.09}_{-  0.17}$&  0.04 &13\\%                pg1307_con_kaspi2000.txt
    PG  1411+442&  0.0896&$ 44.56\pm  0.02$&$  8.28_{-  0.30}^{+  0.17}$&$ -0.23_{-  0.38}^{+  0.63}$&$ 10.51\pm  0.99$&  1.64&$ -0.94^{+  0.04}_{-  0.04}$&$  1.35^{+  0.09}_{-  0.33}$&  0.01 &13\\%                pg1411_con_kaspi2000.txt
    PG  1426+015&  0.0866&$ 44.63\pm  0.02$&$  8.97_{-  0.22}^{+  0.12}$&$ -1.51_{-  0.28}^{+  0.47}$&$ 17.34\pm  1.71$&  2.21&$ -0.54^{+  0.17}_{-  0.17}$&$  2.95^{+  0.37}_{-  0.39}$&  0.36 &13\\%                pg1426_con_kaspi2000.txt
    PG  1613+658&  0.1290&$ 44.77\pm  0.02$&$  8.81_{-  0.21}^{+  0.14}$&$ -0.97_{-  0.31}^{+  0.45}$&$ 12.31\pm  0.92$&  1.63&$ -0.72^{+  0.20}_{-  0.15}$&$  2.95^{+  0.39}_{-  0.37}$&  0.38 &13\\%                pg1613_con_kaspi2000.txt
    PG  1617+175&  0.1124&$ 44.39\pm  0.02$&$  8.79_{-  0.28}^{+  0.15}$&$ -1.50_{-  0.33}^{+  0.58}$&$ 19.12\pm  1.59$&  2.16&$ -0.70^{+  0.13}_{-  0.09}$&$  2.35^{+  0.28}_{-  0.17}$&  0.09 &13\\%                pg1617_con_kaspi2000.txt
    PG  1700+518&  0.2920&$ 45.59\pm  0.01$&$  8.40_{-  0.08}^{+  0.08}$&$  1.08_{-  0.17}^{+  0.17}$&$  6.04\pm  0.50$&  1.28&$ -1.14^{+  0.09}_{-  0.07}$&$  2.11^{+  0.17}_{-  0.22}$&  0.06 &13\\%                pg1700_con_kaspi2000.txt
   Ark       120&  0.0327&$ 43.98\pm  0.06$&$  8.53_{-  0.13}^{+  0.07}$&$ -1.48_{-  0.23}^{+  0.24}$&$  3.95\pm  0.79$&  1.16&$ -1.31^{+  0.28}_{-  0.13}$&$  1.86^{+  0.59}_{-  0.61}$&  0.38 &2\\%               ark120_peterson1998_1.txt
   Ark       120&  0.0327&$ 43.63\pm  0.08$&$  8.45_{-  0.07}^{+  0.05}$&$ -2.01_{-  0.27}^{+  0.27}$&$  8.11\pm  1.27$&  1.34&$ -1.02^{+  0.28}_{-  0.15}$&$  2.09^{+  0.54}_{-  0.46}$&  0.58 &2\\%               ark120_peterson1998_2.txt
   Mrk       110&  0.0353&$ 43.68\pm  0.04$&$  7.05_{-  0.18}^{+  0.09}$&$  0.81_{-  0.32}^{+  0.35}$&$ 10.15\pm  1.65$&  1.60&$ -0.96^{+  0.07}_{-  0.09}$&$  1.02^{+  0.24}_{-  0.98}$&  0.06 &2\\%               mrk110_peterson1998_1.txt
   Mrk       110&  0.0353&$ 43.75\pm  0.04$&$  7.04_{-  0.16}^{+  0.18}$&$  0.92_{-  0.32}^{+  0.34}$&$ 12.59\pm  2.44$&  1.47&$ -0.59^{+  0.28}_{-  0.24}$&$  2.28^{+  0.54}_{-  0.59}$&  1.60 &2\\%               mrk110_peterson1998_2.txt
   Mrk       110&  0.0353&$ 43.53\pm  0.05$&$  7.22_{-  0.16}^{+  0.16}$&$  0.58_{-  0.33}^{+  0.35}$&$ 32.33\pm  4.35$&  2.91&$ -0.25^{+  0.22}_{-  0.22}$&$  3.00^{+  0.41}_{-  0.63}$&  4.14 &2\\%               mrk110_peterson1998_3.txt
   Mrk       279&  0.0305&$ 43.71\pm  0.07$&$  7.97_{-  0.12}^{+  0.09}$&$ -0.89_{-  0.30}^{+  0.33}$&$  9.17\pm  0.81$&  1.65&$ -0.79^{+  0.26}_{-  0.22}$&$  2.33^{+  0.52}_{-  0.52}$&  1.12 &5\\%             mrk279_con_santos2001_1.txt
   Mrk       335&  0.0258&$ 43.76\pm  0.06$&$  7.02_{-  0.12}^{+  0.11}$&$  1.28_{-  0.29}^{+  0.30}$&$ 13.36\pm  0.84$&  1.57&$ -0.91^{+  0.24}_{-  0.13}$&$  1.78^{+  0.52}_{-  0.30}$&  0.45 &1\\%                mrk335_con_grier2012.txt
   Mrk       335&  0.0258&$ 43.84\pm  0.06$&$  6.84_{-  0.25}^{+  0.18}$&$  1.39_{-  0.29}^{+  0.30}$&$  6.59\pm  1.04$&  1.29&$ -0.80^{+  0.26}_{-  0.22}$&$  2.30^{+  0.52}_{-  0.43}$&  1.12 &2\\%               mrk335_peterson1998_1.txt
   Mrk       335&  0.0258&$ 43.74\pm  0.06$&$  6.92_{-  0.14}^{+  0.11}$&$  1.25_{-  0.29}^{+  0.30}$&$  4.90\pm  0.81$&  1.22&$ -1.12^{+  0.28}_{-  0.13}$&$  1.82^{+  0.59}_{-  0.35}$&  0.30 &2\\%               mrk335_peterson1998_2.txt
   Mrk       590&  0.0264&$ 43.59\pm  0.06$&$  7.50_{-  0.06}^{+  0.07}$&$ -0.22_{-  0.25}^{+  0.24}$&$  7.51\pm  1.17$&  1.38&$ -1.25^{+  0.17}_{-  0.09}$&$  1.31^{+  0.46}_{-  0.26}$&  0.09 &2\\%               mrk590_peterson1998_1.txt
   Mrk       590&  0.0264&$ 43.14\pm  0.09$&$  7.58_{-  0.48}^{+  0.22}$&$ -0.91_{-  0.30}^{+  0.28}$&$ 10.34\pm  1.85$&  1.36&$ -0.41^{+  0.22}_{-  0.24}$&$  2.58^{+  0.41}_{-  0.56}$&  1.94 &2\\%               mrk590_peterson1998_2.txt
   Mrk       590&  0.0264&$ 43.38\pm  0.07$&$  7.63_{-  0.09}^{+  0.07}$&$ -0.54_{-  0.26}^{+  0.25}$&$  6.85\pm  1.31$&  1.27&$ -0.83^{+  0.30}_{-  0.13}$&$  2.01^{+  0.59}_{-  0.52}$&  0.67 &2\\%               mrk590_peterson1998_3.txt
   Mrk       590&  0.0264&$ 43.65\pm  0.06$&$  7.55_{-  0.07}^{+  0.05}$&$ -0.13_{-  0.25}^{+  0.24}$&$ 15.34\pm  2.68$&  1.72&$ -1.06^{+  0.26}_{-  0.17}$&$  2.17^{+  0.56}_{-  0.43}$&  0.92 &2\\%               mrk590_peterson1998_4.txt
   Mrk        79&  0.0222&$ 43.63\pm  0.07$&$  7.65_{-  0.88}^{+  0.28}$&$ -0.75_{-  0.34}^{+  0.41}$&$  9.29\pm  1.56$&  1.36&$ -1.14^{+  0.26}_{-  0.13}$&$  1.88^{+  0.52}_{-  0.43}$&  0.38 &2\\%                mrk79_peterson1998_1.txt
   Mrk        79&  0.0222&$ 43.74\pm  0.07$&$  7.85_{-  0.23}^{+  0.15}$&$ -0.59_{-  0.34}^{+  0.41}$&$  9.97\pm  1.68$&  1.44&$ -0.65^{+  0.22}_{-  0.17}$&$  2.27^{+  0.48}_{-  0.48}$&  0.95 &2\\%                mrk79_peterson1998_2.txt
   Mrk        79&  0.0222&$ 43.66\pm  0.07$&$  7.85_{-  0.20}^{+  0.15}$&$ -0.70_{-  0.34}^{+  0.41}$&$  9.54\pm  1.41$&  1.41&$ -0.61^{+  0.22}_{-  0.26}$&$  2.81^{+  0.46}_{-  0.59}$&  2.83 &2\\%                mrk79_peterson1998_3.txt
   Mrk       817&  0.0315&$ 43.79\pm  0.05$&$  7.92_{-  0.09}^{+  0.08}$&$ -0.81_{-  0.35}^{+  0.35}$&$  4.86\pm  0.28$&  1.27&$ -0.93^{+  0.30}_{-  0.13}$&$  2.09^{+  0.54}_{-  0.54}$&  0.84 &6\\%               mrk817_con_denney2010.txt
   Mrk       817&  0.0315&$ 43.67\pm  0.05$&$  7.91_{-  0.11}^{+  0.09}$&$ -0.98_{-  0.35}^{+  0.35}$&$ 13.47\pm  1.96$&  1.59&$ -1.22^{+  0.28}_{-  0.17}$&$  2.07^{+  0.56}_{-  0.52}$&  0.57 &2\\%               mrk817_peterson1998_1.txt
   Mrk       817&  0.0315&$ 43.67\pm  0.05$&$  8.17_{-  0.11}^{+  0.08}$&$ -0.98_{-  0.35}^{+  0.35}$&$  9.61\pm  1.72$&  1.41&$ -0.83^{+  0.22}_{-  0.26}$&$  2.45^{+  0.46}_{-  0.52}$&  2.34 &2\\%               mrk817_peterson1998_2.txt
   Mrk       817&  0.0315&$ 43.84\pm  0.05$&$  7.94_{-  0.12}^{+  0.09}$&$ -0.73_{-  0.35}^{+  0.35}$&$  4.95\pm  0.94$&  1.23&$ -0.51^{+  0.24}_{-  0.24}$&$  2.53^{+  0.50}_{-  0.56}$&  2.05 &2\\%               mrk817_peterson1998_3.txt
   NGC      4051&  0.0023&$ 41.96\pm  0.19$&$  5.42_{-  0.53}^{+  0.23}$&$  1.59_{-  0.84}^{+  1.29}$&$  8.45\pm  0.50$&  1.69&$ -1.05^{+  0.07}_{-  0.04}$&$  0.82^{+  0.15}_{-  0.15}$&  0.05 &14\\%              ngc4051_con_denney2009.txt
    PG  0844+349&  0.0640&$ 44.22\pm  0.07$&$  7.66_{-  0.23}^{+  0.15}$&$  0.50_{-  0.42}^{+  0.57}$&$ 10.49\pm  0.84$&  1.68&$ -0.93^{+  0.04}_{-  0.07}$&$  1.99^{+  0.09}_{-  0.17}$&  0.04 &13\\%                pg0844_con_kaspi2000.txt
    PG  1211+143&  0.0809&$ 44.73\pm  0.08$&$  7.87_{-  0.26}^{+  0.11}$&$  0.84_{-  0.35}^{+  0.63}$&$ 13.44\pm  1.01$&  2.63&$ -0.69^{+  0.13}_{-  0.09}$&$  2.52^{+  0.26}_{-  0.20}$&  0.13 &13\\%                pg1211_con_kaspi2000.txt
   NGC      5273&  0.0036&$ 41.54\pm  0.16$&$  7.14_{-  0.56}^{+  0.19}$&$ -2.50_{-  0.67}^{+  1.33}$&$  7.07\pm  0.93$&  1.34&$ -1.19^{+  0.04}_{-  0.09}$&$  0.05^{+  0.04}_{-  0.22}$&  0.02 &17\\%                           Bentz2014.txt
   MCG    -06-30-15&  0.0077&$ 41.85\pm  0.21$&$  6.60_{-  0.15}^{+  0.15}$&$ -0.94_{-  0.44}^{+  0.44}$&$  6.60\pm  0.58$&  1.34&$ -1.15^{+  0.37}_{-  0.07}$&$  0.86^{+  0.91}_{-  0.13}$&  0.06 &25\\%                              Bentz2016a
   UGC     06728&  0.0065&$ 41.77\pm  0.10$&$  5.55_{-  0.25}^{+  0.22}$&$  1.04_{-  0.72}^{+  0.68}$&$  9.01\pm  1.27$&  1.47&$ -1.03^{+  0.15}_{-  0.07}$&$  0.21^{+  0.48}_{-  0.22}$&  0.04 &26\\%                              Bentz2016b
    3C       382&  0.0579&$ 44.15\pm  0.05$&$  8.01_{-  0.05}^{+  0.09}$&$ -0.31_{-  0.59}^{+  0.61}$&$  8.71\pm  0.45$&  1.44&$ -0.85^{+  0.41}_{-  0.24}$&$  2.66^{+  0.83}_{-  0.39}$&  2.50 &27\\%                           Fausnaugh2017
   MCG   +08-11-011&  0.0205&$ 43.38\pm  0.05$&$  6.61_{-  0.02}^{+  0.02}$&$  1.33_{-  0.08}^{+  0.08}$&$  9.72\pm  0.52$&  1.48&$ -0.84^{+  0.63}_{-  0.15}$&$  1.88^{+  1.26}_{-  0.22}$&  0.49 &27\\%                           Fausnaugh2017
   Mrk       374&  0.0426&$ 43.95\pm  0.05$&$  7.49_{-  0.10}^{+  0.17}$&$  0.45_{-  0.21}^{+  0.34}$&$  2.81\pm  0.26$&  1.23&$ -1.48^{+  0.28}_{-  0.07}$&$  0.93^{+  0.65}_{-  0.13}$&  0.07 &27\\%                           Fausnaugh2017
   NGC      2617&  0.0142&$ 43.12\pm  0.05$&$  7.37_{-  0.14}^{+  0.11}$&$ -0.59_{-  0.28}^{+  0.23}$&$  8.69\pm  0.53$&  1.54&$ -0.89^{+  0.59}_{-  0.15}$&$  1.92^{+  1.09}_{-  0.30}$&  0.54 &27\\%                           Fausnaugh2017
   NGC      4051&  0.0023&$ 41.92\pm  0.04$&$  5.52_{-  0.20}^{+  0.13}$&$  1.35_{-  0.41}^{+  0.27}$&$  2.20\pm  0.12$&  1.13&$ -1.67^{+  0.03}_{-  0.02}$&$  0.70^{+  0.06}_{-  0.04}$&  0.02 &27\\%                           Fausnaugh2017
   NGC      5548&  0.0172&$ 43.44\pm  0.03$&$  7.87_{-  0.05}^{+  0.05}$&$ -1.08_{-  0.12}^{+  0.12}$&$  6.67\pm  0.32$&  1.34&$ -1.11^{+  0.46}_{-  0.11}$&$  1.51^{+  1.00}_{-  0.13}$&  0.16 &28 %                                 Pei2017
\enddata
\tablecomments{
1. Column (1) is the name of RM AGNs, Columns (2$-$5) list the redshift, 
optical luminosity at 5100\AA~corrected for the starlight of host galaxy, black hole mass and accretion rates. 
The model independent variability parameters including traditional variability amplitude ($F_{\rm var}$) and the ratio of maximum to minimum flux ($R_{\rm max}$) in the light curve are listed in Columns (6) and (7). 
The parameters of DRW model were transformed into the rest frame and listed in column (8; $\log \Sigma_{\rm d}$) 
and column (9; $\log \tau_{\rm d}$). 
Column (10) is the ratio of damped variability timescale ($\tau_{\rm d}$) to the observation length of light curves ($D$). 
} 
%the unit of $\langle F\rangle$ is $10^{15} \ergs \rm cm^{-2}$ \AA\ $^{-1}$, and $\sigma_{\rm d}$ is $10^{15} \ergs \rm cm^{-2}$ \AA\ $^{-1}$.\\
\tablecomments{
2. REFERENCES: (1) \citealt{Grier2012}; (2) \citealt{Peterson1998}; 
(3) \citealt{Dietrich1998}; (4) \citealt{Bentz2009}; (5) \citealt{Santos-Lleo1997}; 
(6) \citealt{Denney2010}; (7) \citealt{Stirpe1994}; (8) \citealt{Bentz2006};  
(9) \citealt{Denney2006}; (10) \citealt{Collier1998}; (11) \citealt{Bentz2007}; 
(12) \citealt{Peterson2002}; (13) \citealt{Kaspi2000}; (14) \citealt{Denney2009}; 
(15) \citealt{Barth2011}; (16) \citealt{Dietrich2012}; (17) \citealt{Bentz2014}; 
(18) \citealt{Lu2016a}; (19) \citealt{Du2014};  (20) \citealt{Wang2014}; (21) \citealt{Hu2015}; 
(22) \citealt{Du2015}; (23) \citealt{Du2016}; (24) \citealt{Du2018}; 
(25) \citealt{Bentz2016a}; (26) \citealt{Bentz2016b}; (27) \citealt{Fausnaugh2017}; (28) \citealt{Pei2017}.
} 
\end{deluxetable*}

%\clearpage

\begin{deluxetable*}{llllllll}
\tablecolumns{5}
\tablewidth{0pt}
\setlength{\tabcolsep}{4pt}
\tablecaption{Results of Correlation Analysis \label{tab:pr}}
\tabletypesize{\scriptsize}
\tablehead{
\colhead{}                   &
\colhead{Para.}                   &
\colhead{$\Sigma_{\rm d}$}          &
\colhead{$\tau_{\rm H\beta}$}&
\colhead{$\bhm$}             &
\colhead{$L_{5100}$}         &
\colhead{$\mathdotM$}        
}
\startdata
(1)&$\taud$                    &                                                       & (0.57, $1.43\times10^{-4}$) & (0.68, $2.02\times10^{-6}$)  & (0.31, 0.05)  & (-0.16, 0.34) \\
(2)&$\Sigma_{\rm d}$                &                                                      & (0.19, 0.23)                           & (0.55, $2.75\times10^{-4}$)   & (0.08, 0.65)                            & (-0.33, 0.04) \\
(3)&$F_{\rm var}$          &(0.93, $6.57\times10^{-18}$)         & (0.18, 0.26)                          & (0.54, $3.62\times10^{-4}$)   & (0.11, 0.50)                             & (-0.34, 0.03) \\
(4)$^{*}$&$F_{\rm var}$          &(0.79, $3.64\times10^{-25}$)       & (0.15, 0.10)                         & (0.22, 0.02)                              & (0.04, 0.69)                           & (-0.25, 0.01)                
\enddata
\tablecomments{
Columns (1), (2) and (3) are results of correlation analysis for $\tau_{d}|_{<0.1D}$ sample,
and Column (4)$^{*}$ is result of correlation analysis for all RM AGNs. 
The numbers in `()' are the Pearson correlation coefficient ($\rho$) and null-probability ($p$), respectively.  \\
}
\end{deluxetable*}

\end{document}